\documentclass[modern]{aastex62}
\usepackage{newtxtext,newtxmath} 
\usepackage[USenglish]{babel}
\usepackage[utf8]{inputenc}
\usepackage[T1]{fontenc}
\usepackage{comment}
\usepackage{enumitem}
\usepackage{glossaries}
\usepackage{todonotes}
\usepackage{collect}
\usepackage{wrapfig}

\setlist{noitemsep}

\graphicspath{{./}{./figures/}{./tabs}}

\newcommand{\Contact}[1]{}
\newcommand{\Contributors}[1]{ {\bf Contributors:} \textit{#1} }
\newcommand{\mail}[1]{\href{mailto:#1}{#1}}

\newcommand{\Comment}[3]{\textcolor{#1}{(#2: #3)}}
\newcommand{\WOM}[1]{\Comment{red}{WOM}{#1}} 

\providecommand{\secref}[1]{\hyperref[#1]{Section~\ref{#1}}}

\providecommand{\figref}[1]{Figure~\ref{#1}}

\providecommand{\recref}[1]{\hyperref[#1]{REC-\ref{#1}}}

\newcounter{reccount} 

\newenvironment{recenv}[1]
    { 
    \refstepcounter{reccount}  
    #1
    }

\definecollection{nrecommendations}

\newcommand{\nrec}[6]{ 
	\begin{recenv}
	\vspace{5pt}
	\noindent \textbf{REC-\thereccount~#5}. \newline
	\textbf{Area:} #1.  \textbf{Audience:} #2. \textbf {Term:} #3
	\newline \label{rec:#4}\noindent \textit{#6}
	\vspace{5pt}
	\end{recenv}
    \begin{collect}{nrecommendations}{}{}
        \noindent \textbf{REC-\ref{rec:#4}~#5}.
	\newline \textbf{Area:} #1.  \textbf{Audience:} #2. \textbf {Term:} #3
        \newline \textit{#6}.
        \newline See page \pageref{rec:#4}. \vspace{10pt}
     \end{collect}
\typeout{REC-\thereccount: #1: #2: #3: #4: #5 :END}
}

\makeatletter
\renewcommand{\section}{\newpage{} \@startsection
{section}
{1}
{0mm}
{-3.5ex \@plus -1ex \@minus -.2ex}
{2.3ex \@plus.2ex}
{\normalfont\bfseries}}
\makeatother

\newglossaryentry{1D} {name={1D}, description={One-dimensional}}
\newglossaryentry{2D} {name={2D}, description={Two-dimensional}}
\newglossaryentry{3D} {name={3D}, description={Three-dimensional}}
\newglossaryentry{AAG} {name={AAG}, description={Astronomy and Astrophysics Research Grants, an \gls{NSF} program}}
\newacronym{AAS} {AAS} {American Astronomical Society}
\newacronym{ADS} {ADS} {Astrophysics Data System}
\newglossaryentry{AI} {name={AI}, description={Artificial Intelligence, used to cover many machine learning algorithms}}
\newglossaryentry{AJ} {name={AJ}, description={The Astronomical Journal}}
\newglossaryentry{APC} {name={APC}, description={activities, projects, or state of the profession considerations - wrt. the decadal survey}}
\newglossaryentry{API} {name={API}, description={Application Programming Interface.  Usually this is either "web API" (meaning how you access/update data via http) or "software API" (meaning how you call the function/class/etc in a particular programming language or library), although often only "API" is used and the prefix is implicit.}}
\newacronym{APL} {APL} {Apache Public License}
\newglossaryentry{ATLAS} {name={ATLAS}, description={The Asteroid Terrestrial-impact Last Alert System}}
\newglossaryentry{AURA} {name={AURA}, description={\gls{Association of Universities for Research in Astronomy}}}
\newglossaryentry{Alert} {name={Alert}, description={A packet of information for each source detected with signal-to-noise ratio > 5 in a difference image during Prompt Processing, containing measurement and characterization parameters based on the past 12 months of LSST observations plus small cutouts of the single-visit, template, and difference images, distributed via the internet}}
\newglossaryentry{Alert Production} {name={Alert Production}, description={The principal component of Prompt Processing that processes and calibrates incoming images, performs Difference Image Analysis to identify DIASources and DIAObjects, packages and distributes the resulting Alerts, and runs the Moving Object Processing System}}
\newglossaryentry{Apache Parquet} {name={Apache Parquet}, description={A columnar storage data persistence format maintained by the Apache project}}
\newglossaryentry{Archive} {name={Archive}, description={The repository for documents required by the NSF to be kept. These include documents related to design and development, construction, integration, test, and operations of the LSST observatory system. The archive is maintained using the enterprise content management system DocuShare, which is accessible through a link on the project website www.project.lsst.org}}
\newglossaryentry{Archive Center} {name={Archive Center}, description={Part of the LSST Data Management System, the LSST archive center is a data center at NCSA that hosts the LSST Archive, which includes released science data and metadata, observatory and engineering data, and supporting software such as the LSST Software Stack}}
\newglossaryentry{Association Pipeline} {name={Association Pipeline}, description={An application that matches detected Sources or DIASources or generated Objects to an existing catalog of Objects, producing a (possibly many-to-many) set of associations and a list of unassociated inputs. Association Pipelines are used in Prompt Processing after DIASource generation and in the final stages of Data Release processing to ensure continuity of Object identifiers}}
\newglossaryentry{Association of Universities for Research in Astronomy} {name={Association of Universities for Research in Astronomy}, description={ consortium of US institutions and international affiliates that operates world-class astronomical observatories, AURA is the legal entity responsible for managing what it calls independent operating Centers, including LSST, under respective cooperative agreements with the National Science Foundation. AURA assumes fiducial responsibility for the funds provided through those cooperative agreements. AURA also is the legal owner of the AURA Observatory properties in Chile}}
\newglossaryentry{B} {name={B}, description={Byte (8 bit)}}
\newglossaryentry{Butler} {name={Butler}, description={A middleware component for persisting and retrieving image datasets (raw or processed), calibration reference data, and catalogs}}
\newacronym{CADC} {CADC} {Canadian Astronomy Data Centre}
\newglossaryentry{CAOM} {name={CAOM}, description={Common Archive Observation Model \url{http://www.opencadc.org/caom2/}}}
\newglossaryentry{CCD} {name={CCD}, description={\gls{Charge-Coupled Device}}}
\newglossaryentry{CI} {name={CI}, description={\gls{cyberinfrastructure}}}
\newacronym{CMB} {CMB} {Cosmic Microwave Background}
\newacronym{CPU} {CPU} {Central Processing Unit}
\newacronym{CS} {CS} {Computer Science}
\newglossaryentry{CSSI} {name={CSSI}, description={Cyberinfrastructure for Sustained Scientific Innovation  \url{https://www.nsf.gov/pubs/2019/nsf19548/nsf19548.htm}}}
\newglossaryentry{Center} {name={Center}, description={An entity managed by AURA that is responsible for execution of a federally funded project}}
\newglossaryentry{Charge-Coupled Device} {name={Charge-Coupled Device}, description={a particular kind of solid-state sensor for detecting optical-band photons. It is composed of a 2-D array of pixels, and one or more read-out amplifiers}}
\newglossaryentry{Citizen Science} {name={Citizen Science}, description={- Amanda, Arfon, Meg?,}}
\newglossaryentry{Construction} {name={Construction}, description={The period during which LSST observatory facilities, components, hardware, and software are built, tested, integrated, and commissioned. Construction follows design and development and precedes operations. The LSST construction phase is funded through the \gls{NSF} \gls{MREFC} account}}
\newglossaryentry{DB} {name={DB}, description={DataBase}}
\newacronym{DCR} {DCR} {Differential Chromatic Refraction}
\newacronym{DES} {DES} {Dark Energy Survey}
\newacronym{DESI} {DESI} {Dark Energy Spectroscopic Instrument}
\newacronym{DIA} {DIA} {Difference Image Analysis}
\newglossaryentry{DIAObject} {name={DIAObject}, description={A DIAObject is the association of DIASources, by coordinate, that have been detected with signal-to-noise ratio greater than 5 in at least one difference image. It is distinguished from a regular Object in that its brightness varies in time, and from a SSObject in that it is stationary (non-moving)}}
\newglossaryentry{DIASource} {name={DIASource}, description={A DIASource is a detection with signal-to-noise ratio greater than 5 in a difference image}}
\newglossaryentry{DIBBs} {name={DIBBs}, description={Data Infrastructure Building Blocks}}
\newglossaryentry{DKIST} {name={DKIST}, description={Daniel K. Inouye Solar Telescope (formerly the Advanced Technology Solar Telescope, ATST)}}
\newacronym{DM} {DM} {\gls{Data Management}}
\newacronym{DMS} {DMS} {Data Management Subsystem}
\newacronym{DOE} {DOE} {\gls{Department of Energy}}
\newglossaryentry{DOI} {name={DOI}, description={Digital Object Identifier \url{https://www.doi.org/}}}
\newacronym{DR} {DR} {Data Release}
\newacronym{DRP} {DRP} {Data Release Production}
\newglossaryentry{DVCS} {name={DVCS}, description={Distributed Version Control System, a form of \gls{version control} where the complete codebase - including its full history - is mirrored on every developer's computer}}
\newglossaryentry{Data Management} {name={Data Management}, description={The LSST Subsystem responsible for the Data Management System (DMS), which will capture, store, catalog, and serve the LSST dataset to the scientific community and public. The DM team is responsible for the DMS architecture, applications, middleware, infrastructure, algorithms, and Observatory Network Design. DM is a distributed team working at LSST and partner institutions, with the DM Subsystem Manager located at LSST headquarters in Tucson}}
\newglossaryentry{Data Management Subsystem} {name={Data Management Subsystem}, description={The Data Management Subsystem is one of the four subsystems which constitute the LSST Construction Project. The Data Management Subsystem is responsible for developing and delivering the LSST Data Management System to the LSST Operations Project}}
\newglossaryentry{Data Management System} {name={Data Management System}, description={The computing infrastructure, middleware, and applications that process, store, and enable information extraction from the LSST dataset; the DMS will process peta-scale data volume, convert raw images into a faithful representation of the universe, and archive the results in a useful form. The infrastructure layer consists of the computing, storage, networking hardware, and system software. The middleware layer handles distributed processing, data access, user interface, and system operations services. The applications layer includes the data pipelines and the science data archives' products and services}}
\newglossaryentry{Data Release} {name={Data Release}, description={The approximately annual reprocessing of all LSST data, and the installation of the resulting data products in the LSST Data Access Centers, which marks the start of the two-year proprietary period}}
\newglossaryentry{Data Release Production} {name={Data Release Production}, description={An episode of (re)processing all of the accumulated LSST images, during which all output DR data products are generated. These episodes are planned to occur annually during the LSST survey, and the processing will be executed at the Archive Center. This includes Difference Imaging Analysis, generating deep Coadd Images, Source detection and association, creating Object and Solar System Object catalogs, and related metadata}}
\newglossaryentry{Department of Energy} {name={Department of Energy}, description={cabinet department of the United States federal government; the DOE has assumed technical and financial responsibility for providing the LSST camera. The DOE's responsibilities are executed by a collaboration led by SLAC National Accelerator Laboratory}}
\newglossaryentry{Difference Image} {name={Difference Image}, description={Refers to the result formed from the pixel-by-pixel difference of two images of the sky, after warping to the same pixel grid, scaling to the same photometric response, matching to the same PSF shape, and applying a correction for Differential Chromatic Refraction. The pixels in a difference thus formed should be zero (apart from noise) except for sources that are new, or have changed in brightness or position. In the LSST context, the difference is generally taken between a visit image and template. }}
\newglossaryentry{Difference Image Analysis} {name={Difference Image Analysis}, description={The detection and characterization of sources in the Difference Image that are above a configurable threshold, done as part of Alert Generation Pipeline}}
\newglossaryentry{Differential Chromatic Refraction} {name={Differential Chromatic Refraction}, description={The refraction of incident light by Earth's atmosphere causes the apparent position of objects to be shifted, and the size of this shift depends on both the wavelength of the source and its airmass at the time of observation. DCR corrections are done as a part of DIA}}
\newglossaryentry{Docker} {name={Docker}, description={A popular implementation of \gls{container} technology.}}
\newglossaryentry{DocuShare} {name={DocuShare}, description={The trade name for the enterprise management software used by LSST to archive and manage documents}}
\newacronym{ELT} {ELT} {Extremely Large Telescope}
\newglossaryentry{EPO} {name={EPO}, description={Education and Public Outreach}}
\newacronym{ESA} {ESA} {European Space Agency}
\newacronym{ESAC} {ESAC} {European Space Astronomy Centre}
\newacronym{EU} {EU} {European Union}
\newglossaryentry{Education and Public Outreach} {name={Education and Public Outreach}, description={The LSST subsystem responsible for the cyberinfrastructure, user interfaces, and outreach programs necessary to connect educators, planetaria, citizen scientists, amateur astronomers, and the general public to the transformative LSST dataset}}
\newglossaryentry{FITS} {name={FITS}, description={Flexible Image Transport System,  is an open standard defining a digital file format useful for storage, transmission and processing of data. Files are formatted as tables or 2D images}}
\newglossaryentry{FORCE11} {name={FORCE11}, description={FORCE11 is a community of scholars, librarians, archivists, publishers and research funders interested in the Future of Research Communications and e-Scholarship}}
\newglossaryentry{FPGA} {name={FPGA}, description={Field Programmable Gate Array, an integrated circuit which is fairly easily configurable.}}
\newglossaryentry{Flexible Image Transport System} {name={Flexible Image Transport System}, description={an international standard in astronomy for storing images, tables, and metadata in disk files. See the IAU FITS Standard for details}}
\newglossaryentry{GB} {name={GB}, description={Gigabyte}}
\newglossaryentry{GNU} {name={GNU}, description={GNU's Not Unix! An operating system and an extensive collection of free computer software}}
\newacronym{GPL} {GPL} {GNU Public License}
\newacronym{GW} {GW} {Gravity Wave}
\newacronym{HPC} {HPC} {High Performance Computing}
\newacronym{HTC} {HTC} {High Throughput  Computing}
\newglossaryentry{IAM} {name={IAM}, description={Identity and Access Management}}
\newacronym{IAU} {IAU} {International Astronomical Union}
\newglossaryentry{IDL} {name={IDL}, description={Interactive Data Language, a programming language used for data analysis. Harris Geospatial \footnote{\url{https://www.harrisgeospatial.com/Software-Technology/IDL}}}}
\newglossaryentry{IPAC} {name={IPAC}, description={No longer an acronym; science and data center at Caltech}}
\newacronym{IR} {IR} {Infra Red}
\newglossaryentry{IRAF} {name={IRAF}, description={Image Reduction and Analysis Facility,  a collection of software written at the National Optical Astronomy Observatory (NOAO) geared towards the reduction of astronomical images in pixel array form.}}
\newglossaryentry{IRSA} {name={IRSA}, description={Infrared Science Archive}}
\newglossaryentry{ITAR} {name={ITAR}, description={International Traffic in Arms Regulations}}
\newacronym{IUSE} {IUSE} {Improving Undergraduate STEM Education}
\newglossaryentry{IVOA} {name={IVOA}, description={International Virtual-Observatory Alliance}}
\newglossaryentry{JWST} {name={JWST}, description={James Webb Space Telescope (formerly known as NGST)}}
\newglossaryentry{LIGO} {name={LIGO}, description={The Laser Interferometer Gravitational-Wave Observatory}}
\newglossaryentry{LISA} {name={LISA}, description={Laser Interferometer Space Antenna - ESA mission for 2030's}}
\newacronym{LSST} {LSST} {Large Synoptic Survey Telescope}
\newglossaryentry{LSSTC} {name={LSSTC}, description={LSST Corporation, a not for profit organisation associated with LSST}}
\newglossaryentry{LUVOIR} {name={LUVOIR}, description={The Large UV/Optical/IR Surveyor - concept mission.}}
\newglossaryentry{MAST} {name={MAST}, description={Mikulski Archive for Space Telescopes}}
\newglossaryentry{MCMC} {name={MCMC}, description={Markov Chain Monte Carlo, a class of algorithms for sampling from a probability distribution.}}
\newglossaryentry{ML} {name={ML}, description={Machine Learning (see also \gls{AI})}}
\newacronym{MOPS} {MOPS} {Moving Object Processing System}
\newglossaryentry{MREFC} {name={MREFC}, description={\gls{Major Research Equipment and Facility Construction}}}
\newacronym{MSE} {MSE} {Maunakea Spectroscopic Explorer}
\newglossaryentry{Major Research Equipment and Facility Construction} {name={Major Research Equipment and Facility Construction}, description={the NSF account through which large facilities construction projects such as LSST are funded}}
\newglossaryentry{Mapper} {name={Mapper}, description={A piece of software that abstracts persisting and unpersisting data; specifically, it knows how to navigate a data repository to locate data that match selection criteria that are relevant for data obtained with a particular camera. Used by the Butler}}
\newglossaryentry{Moving Object Processing System} {name={Moving Object Processing System}, description={The Moving Object Processing System (MOPS) identifies new SSObjects using unassociated DIASources. MOPS is part of the Science Pipelines}}
\newacronym{NAS} {NAS} {Neural Architecture Search}
\newglossaryentry{NASA} {name={NASA}, description={National Aeronautics and Space Administration}}
\newglossaryentry{NASA ROSES} {name={NASA ROSES}, description={Research Opportunities in Earth and Space Science}}
\newglossaryentry{NCSA} {name={NCSA}, description={National Center for Supercomputing Applications}}
\newglossaryentry{NGSS} {name={NGSS}, description={Next Generation Science Standards \url{https://www.nextgenscience.org/}}}
\newglossaryentry{NGVLA} {name={NGVLA}, description={Next Generation Very Large Array \url{https://science.nrao.edu/futures/ngvla}}}
\newacronym{NIR} {NIR} {Near Infra Red}
\newglossaryentry{NOAO} {name={NOAO}, description={National Optical Astronomy Observatories (USA)}}
\newacronym{NSF} {NSF} {National Science Foundation}
\newacronym{NSTA} {NSTA} {National Science Teachers Association}
\newacronym{NYU} {NYU} {New York University}
\newglossaryentry{National Science Foundation} {name={National Science Foundation}, description={primary federal agency supporting research in all fields of fundamental science and engineering; NSF selects and funds projects through competitive, merit-based review}}
\newglossaryentry{ODBC} {name={ODBC}, description={Open DataBase Connectivity, a standard \gls{API} for \gls{SQL} databases.}}
\newacronym{OS} {OS} {Operating System}
\newglossaryentry{Object} {name={Object}, description={In LSST nomenclature this refers to an astronomical object, such as a star, galaxy, or other physical entity. E.g., comets, asteroids are also Objects but typically called a Moving Object or a Solar System Object (SSObject). One of the DRP data products is a table of Objects detected by LSST which can be static, or change brightness or position with time}}
\newglossaryentry{OpenEXR} {name={OpenEXR}, description={a high dynamic range raster file format, released as an open standard along with a set of software tools created by Industrial Light \& Magic (ILM) \url{http://www.openexr.com/index.html}}}
\newglossaryentry{Operations} {name={Operations}, description={The 10-year period following construction and commissioning during which the LSST Observatory conducts its survey}}
\newglossaryentry{PASP} {name={PASP}, description={Publications of the Astronomical Society of the Pacific}}
\newglossaryentry{PB} {name={PB}, description={PetaByte}}
\newacronym{PDF} {PDF} {Probability Density Function}
\newglossaryentry{PHP} {name={PHP}, description={a popular general-purpose scripting language that is especially suited to web development.}}
\newacronym{PI} {PI} {Principle Investigator}
\newglossaryentry{PLATO} {name={PLATO}, description={PLAnetary Transits and Oscillations of stars, the third medium-class mission in ESA's Cosmic Vision programme.}}
\newglossaryentry{PSF} {name={PSF}, description={Point Spread Function,  describes the response of an imaging system to a point source or point object.}}
\newglossaryentry{Pan-STARRS} {name={Pan-STARRS}, description={Panoramic Survey Telescope and Rapid Response System}}
\newglossaryentry{Project Manager} {name={Project Manager}, description={The person responsible for exercising leadership and oversight over the entire LSST project; he or she controls schedule, budget, and all contingency funds}}
\newglossaryentry{Prompt Processing} {name={Prompt Processing}, description={The processing that occurs at the Archive Center on the nightly stream of raw images coming from the telescope, including Difference Imaging Analysis, Alert Production, and the Moving Object Processing System. This processing generates Prompt Data Products}}
\newacronym{RA} {RA} {Right Ascension}
\newglossaryentry{S3} {name={S3}, description={Structured, imperative high level computer programming language, used as implementation language for the Virtual Machine Environment (\gls{VME}) operating system}}
\newacronym{SDSS} {SDSS} {Sloan Digital Sky Survey}
\newacronym{SKA} {SKA} {Square Kilometer Array}
\newglossaryentry{SLAC} {name={SLAC}, description={SLAC National Accelerator Laboratory (formerly Stanford Linear Accelerator Center; SLAC is now no longer an acronym)}}
\newglossaryentry{SQL} {name={SQL}, description={Structured Query Language, for interrogating relation databases}}
\newglossaryentry{STEM} {name={STEM}, description={Science, Technology, Engineering and Math}}
\newglossaryentry{Science Pipelines} {name={Science Pipelines}, description={The library of software components and the algorithms and processing pipelines assembled from them that are being developed by DM to generate science-ready data products from LSST images. The Pipelines may be executed at scale as part of LSST Prompt or Data Release processing, or pieces of them may be used in a standalone mode or executed through the LSST Science Platform. The Science Pipelines are one component of the LSST Software Stack}}
\newglossaryentry{Science Platform} {name={Science Platform}, description={A set of integrated web applications and services deployed at the LSST Data Access Centers (DACs) through which the scientific community will access, visualize, and perform next-to-the-data analysis of the LSST data products}}
\newglossaryentry{Sloan Digital Sky Survey} {name={Sloan Digital Sky Survey}, description={is a digital survey of roughly 10,000 square degrees of sky around the north Galactic pole, plus a ~300 square degree stripe along the celestial equator}}
\newglossaryentry{Software Stack} {name={Software Stack}, description={Often referred to as the LSST Stack, or just The Stack, it is the collection of software written by the LSST Data Management Team to process, generate, and serve LSST images, transient alerts, and catalogs. The Stack includes the LSST Science Pipelines, as well as packages upon which the DM software depends. It is open source and publicly available}}
\newglossaryentry{Solar System Object} {name={Solar System Object}, description={A solar system object is an astrophysical object that is identified as part of the Solar System: planets and their satellites, asteroids, comets, etc. This class of object had historically been referred to within the LSST Project as Moving Objects}}
\newglossaryentry{Source} {name={Source}, description={A single detection of an astrophysical object in an image, the characteristics for which are stored in the Source Catalog of the DRP database. The association of Sources that are non-moving lead to Objects; the association of moving Sources leads to Solar System Objects. (Note that in non-LSST usage "source" is often used for what LSST calls an Object.)}}
\newglossaryentry{Subsystem} {name={Subsystem}, description={A set of elements comprising a system within the larger LSST system that is responsible for a key technical deliverable of the project}}
\newglossaryentry{Subsystem Manager} {name={Subsystem Manager}, description={responsible manager for an LSST subsystem; he or she exercises authority, within prescribed limits and under scrutiny of the Project Manager, over the relevant subsystem's cost, schedule, and work plans}}
\newacronym{TAP} {TAP} {Table Access Protocol}
\newglossaryentry{TB} {name={TB}, description={TeraByte}}
\newglossaryentry{TESS} {name={TESS}, description={Transiting Exoplanet Survey Satellite,  a space telescope for NASA's Explorer program}}
\newacronym{TMT} {TMT} {Thirty Meter Telescope}
\newglossaryentry{TPU} {name={TPU}, description={Tensor Processing Unit , a proprietary type of processor designed by Google in 2016 for use with neural networks and in machine learning projects}}
\newglossaryentry{UA} {name={UA}, description={University of Arizona}}
\newacronym{UK} {UK} {United Kingdom}
\newacronym{US} {US} {United States}
\newacronym{UV} {UV} {Ultra Violet}
\newacronym{VME} {VME} {Virtual Machine Environment}
\newacronym{VO} {VO} {Virtual Observatory}
\newacronym{WCS} {WCS} {\gls{World Coordinate System}}
\newglossaryentry{WFIRST} {name={WFIRST}, description={Wide Field Infrared Survey Telescope}}
\newglossaryentry{World Coordinate System} {name={World Coordinate System}, description={a mapping from image pixel coordinates to physical coordinates; in the case of images the mapping is to sky coordinates, generally in an equatorial (RA, Dec) system. The \gls{WCS} is expressed in FITS file extensions as a collection of header keyword=value pairs (basically, the values of parameters for a selected functional representation of the mapping) that are specified in the FITS Standard}}
\newacronym{ZTF} {ZTF} {Zwicky Transient Facility}
\newglossaryentry{aggregate metric} {name={aggregate metric}, description={An aggregation of multiple point metrics. For example, the overall photometric repeatability for a particular tract given given the repeatability of multiple individual stars in the tract. See also: “metric”}}
\newglossaryentry{aggregation} {name={aggregation}, description={The process of reducing multiple input values to a single output, e.g., a metric value, computed from a collection of input values. For example, a sum or average of a metric computed over patches to produce an aggregate metric at tract level. See also: “metric”, “aggregate metric”}}
\newglossaryentry{airmass} {name={airmass}, description={The pathlength of light from an astrophysical source through the Earth's atmosphere. It is given approximately by sec z, where z is the angular distance from the zenith (the point directly overhead, where airmass = 1.0) to the source}}
\newglossaryentry{algorithm} {name={algorithm}, description={A computational implementation of a calculation or some method of processing}}
\newglossaryentry{astrometry} {name={astrometry}, description={In astronomy, the sub-discipline of astrometry concerns precision measurement of positions (at a reference epoch), and real and apparent motions of astrophysical objects. Real motion means 3-D motions of the object with respect to an inertial reference frame; apparent motions are an artifact of the motion of the Earth. Astrometry per se is sometimes confused with the act of determining a World Coordinate System (WCS), which is a functional characterization of the mapping from pixels in an image or spectrum to world coordinate such as (RA, Dec) or wavelength}}
\newglossaryentry{astronomical object} {name={astronomical object}, description={A star, galaxy, asteroid, or other physical object of astronomical interest. Beware: in non-LSST usage, these are often known as sources}}
\newglossaryentry{background} {name={background}, description={In an image, the background consists of contributions from the sky (e.g., clouds or scattered moonlight), and from the telescope and camera optics, which must be distinguished from the astrophysical background. The sky and instrumental backgrounds are characterized and removed by the LSST processing software using a low-order spatial function whose coefficients are recorded in the image metadata}}
\newglossaryentry{calibration} {name={calibration}, description={The process of translating signals produced by a measuring instrument such as a telescope and camera into physical units such as flux, which are used for scientific analysis. Calibration removes most of the contributions to the signal from environmental and instrumental factors, such that only the astronomical component remains}}
\newglossaryentry{camera} {name={camera}, description={An imaging device mounted at a telescope focal plane, composed of optics, a shutter, a set of filters, and one or more sensors arranged in a focal plane array}}
\newglossaryentry{cloud} {name={cloud}, description={A visible mass of condensed water vapor floating in the atmosphere, typically high above the ground or in interstellar space acting as the birthplace for stars.  Also a way of computing (on other peoples computers leveraging their services and availability).}}
\newglossaryentry{cold storage} {name={cold storage}, description={Data moved to cold storage means data moved to cheaper slower storage such as tape. The assumption is this is no longer accessed frequently.}}
\newglossaryentry{community software} {name={community software}, description={Software developed for and shared among a large group of relatively like-minded users (e.g. astronomers). Typically, but not necessarily, \gls{open source software} and \gls{open development}-based.}}
\newglossaryentry{container} {name={container}, description={A container is a software package that contains everything the software needs to run. This includes the executable program as well as system tools, libraries, and settings. ... For example, a container that includes PHP and MySQL can run identically on both a Linux computer and a Windows machine.}}
\newglossaryentry{cyberinfrastructure} {name={cyberinfrastructure}, description={Sometimes denoted CI, A term first used by the US National Science Foundation (\gls{NSF}), and it typically is used to refer to information technology systems that provide particularly powerful and advanced capabilities.}}
\newglossaryentry{data collection} {name={data collection}, description={A data collection in the second-generation (Gen2) Butler (referred to as a data repository in earlier generations) consists of hierarchically organized data files, an inventory or registry of the contents (i.e., metadata from the data files) stored in an sqlite3 file, and a Mapper file that specifies to the LSST Stack software the camera model to apply when accessing the data in the data repository}}
\newglossaryentry{data repository} {name={data repository}, description={A data repository consists of hierarchically organized data files, an inventory or registry of the contents (i.e., metadata from the data files) stored in an sqlite3 file, and a Mapper file that specifies to the LSST Stack software the camera model to apply when accessing the data in the repository. With the second-generation (Gen2) Butler, the term repository will be replaced by data collection}}
\newglossaryentry{epoch} {name={epoch}, description={Sky coordinate reference frame, e.g., J2000. Alternatively refers to a single observation (usually photometric, can be multi-band) of a variable source}}
\newglossaryentry{flux} {name={flux}, description={Shorthand for radiative flux, it is a measure of the transport of radiant energy per unit area per unit time. In astronomy this is usually expressed in cgs units: erg/cm2/s}}
\newglossaryentry{git} {name={git}, description={The most widely used \gls{DVCS} software}}
\newglossaryentry{interoperability} {name={interoperability}, description={the ability of systems or software to exchange and make use of information between them.}}
\newglossaryentry{metadata} {name={metadata}, description={General term for data about data, e.g., attributes of astronomical objects (e.g. images, sources, astroObjects, etc.) that are characteristics of the objects themselves, and facilitate the organization, preservation, and query of data sets. (E.g., a FITS header contains metadata)}}
\newglossaryentry{metric} {name={metric}, description={A measurable quantity which may be tracked. A metric has a name, description, unit, references, and tags (which are used for grouping). A metric is a scalar by definition. See also: aggregate metric, model metric, point metric}}
\newglossaryentry{metric value} {name={metric value}, description={The result of computing a particular metric on some given data. Note that metric values are typically computed rather than measured. See also: metric}}
\newglossaryentry{model metric} {name={model metric}, description={A metric describing a model related to the data. For example, the coefficients of a 2D polynomial fit to the background of a single CCD exposure}}
\newglossaryentry{open development} {name={open development}, description={A process for developing software that emphasizes all code contribution and decision-making be done in the open, available to as wide a group as possible (This usually means anyone with internet access).}}
\newglossaryentry{open source software} {name={open source software}, description={Open source software is a type of software in which source code is released under a license in which the copyright holder grants users the rights to study, change, and distribute the software to anyone and for any purpose. Note that this is \emph{not} necessarily the same as open to contribution (see \gls{open development}).}}
\newglossaryentry{parquet} {name={parquet}, description={Parquet File Format Hadoop. Parquet, an open source file format for Hadoop. Parquet stores nested data structures in a flat columnar format. Compared to a traditional approach where data is stored in row-oriented approach, parquet is more efficient in terms of storage and performance.}}
\newglossaryentry{pipeline} {name={pipeline}, description={A configured sequence of software tasks (Stages) to process data and generate data products. Example: Association Pipeline}}
\newglossaryentry{point metric} {name={point metric}, description={A metric that is associated with a single entry in a catalog. Examples include the shape of a source, the standard deviation of the flux of an object detected on a Coadd, the flux of an source detected on a difference image}}
\newglossaryentry{provenance} {name={provenance}, description={Information about how LSST images, Sources, and Objects were created (e.g., versions of pipelines, algorithmic components, or templates) and how to recreate them}}
\newglossaryentry{reproducibility} {name={reproducibility}, description={(this one should have many definitions and we have to say WHICH version we are talking about) The ability to combine the same code and data and get the same result, or the ability to use the same code with different data to enforce a result, or there may be others}}
\newglossaryentry{shape} {name={shape}, description={In reference to a Source or Object, the shape is a functional characterization of its spatial intensity distribution, and the integral of the shape is the flux. Shape characterizations are a data product in the DIASource, DIAObject, Source, and Object catalogs}}
\newglossaryentry{sky map} {name={sky map}, description={A sky tessellation for LSST. The Stack includes software to define a geometric mapping from the representation of World Coordinates in input images to the LSST sky map. This tessellation is comprised of individual tracts which are, in turn, comprised of patches}}
\newglossaryentry{software} {name={software}, description={The programs and other operating information used by a computer.}}
\newglossaryentry{sqlite3} {name={sqlite3}, description={A software package external to DM, sqlite3 provides a SQL interface compliant with the DB-API 2.0 specification for SQLite, a self-contained public-domain SQL database engine}}
\newglossaryentry{stack} {name={stack}, description={a grouping, usually in layers (hence stack), of software packages and services to achieve a common goal. Often providing a higher level set of end user oriented services and tools}}
\newglossaryentry{tract} {name={tract}, description={A portion of sky, a spherical convex polygon, within the LSST all-sky tessellation (sky map). Each tract is subdivided into sky patches}}
\newglossaryentry{transient} {name={transient}, description={A transient source is one that has been detected on a difference image, but has not been associated with either an astronomical object or a solar system body}}
\newglossaryentry{version control} {name={version control}, description={The management of changes to documents, computer programs, and other collections of information. Changes are usually identified by a number or letter code. Each revision is associated with a timestamp and the person making the change. Revisions can be compared, restored, and merged}}

\makenoidxglossaries

\begin{document}

\title{\Large Petabytes to Science}


\author{Amanda~E.~Bauer}
\affiliation{Large Synoptic Survey Telescope (LSST/AURA)}
\author{Eric C.~Bellm}
\affiliation{LSST}
\affiliation{DIRAC Institute, Department of Astronomy, University of Washington}
\author{Adam~S.~Bolton}
\affiliation{NOAO}
\author{Surajit~Chaudhuri}
\affiliation{Microsoft Research}
\author{A.J.~Connolly}
\affiliation{DIRAC Institute, Department of Astronomy, University of Washington}
\author{Kelle~L.~Cruz}
\affiliation{Hunter College, City University of New York}
\affiliation{American Museum of Natural History}
\affiliation{Center for Computational Astrophysics, Flatiron Institute}
\author{Vandana~Desai}
\affiliation{Caltech/IPAC}
\author{Alex~Drlica-Wagner}
\affiliation{Fermi National Accelerator Laboratory}
\affiliation{Kavli Institute of Cosmological Physics, University of Chicago}
\author{Frossie~Economou}
\affiliation{Large Synoptic Survey Telescope (LSST/AURA)}
\author{Niall~Gaffney}
\affiliation{Texas Advanced Computing Center}
\author{J.~Kavelaars}
\affiliation{National Research Council of Canada}
\author{J.~Kinney}
\affiliation{Google Inc.}
\author{Ting~S.~Li}
\affiliation{Fermi National Accelerator Laboratory}
\affiliation{Kavli Institute of Cosmological Physics, University of Chicago}
\author{B.~Lundgren}
\affiliation{University of North Carolina Asheville}
\author{R.~Margutti}
\affiliation{Northwestern University}
\author{G.~Narayan}
\affiliation{Space Telescope Science Institute}
\author{B.~Nord}
\affiliation{Fermi National Accelerator Laboratory}
\affiliation{Kavli Institute of Cosmological Physics, University of Chicago}
\affiliation{Department of Astronomy and Astrophysics, University of Chicago}
\author{Dara~J.~Norman}
\affiliation{NOAO}
\author{W.~O'Mullane.}
\affiliation{Large Synoptic Survey Telescope (LSST/AURA)}
\author{S.~Padhi}
\affiliation{Amazon Web Services}
\author{J.~E.~G.~Peek}
\affiliation{Space Telescope Science Institute}
\affiliation{Department of Physics \& Astronomy, The Johns Hopkins University}
\author{C.~Schafer}
\affiliation{Carnegie Mellon University}
\author{Megan ~E.~Schwamb}
\affiliation{Gemini Observatory}
\author{Arfon~M.~Smith}
\affiliation{Space Telescope Science Institute}
\author{Alexander~S.~Szalay}
\affiliation{Department of Computer Science, The Johns Hopkins University}
\affiliation{Department of Physics \& Astronomy, The Johns Hopkins University}
\author{Erik~J.~Tollerud}
\affiliation{Space Telescope Science Institute}
\author{Anne-Marie~Weijmans}
\affiliation{School of Physics and Astronomy, University of St Andrews}


\keywords{Astronomy, Astrophysics, Work Force, Diversity, Inclusion, Software, Algorithms, Data Management, Computing, \gls{HPC}, \gls{HTC}, Networking, Machine Learning, Cloud, Education, Management, Outreach, Workforce  }

\begin{abstract}
A Kavli foundation sponsored workshop on the theme \emph{Petabytes to Science} was held 12$^{th}$ to 14$^{th}$ of February 2019 in Las Vegas.
The aim of the this workshop was to discuss important trends and technologies which may support astronomy. We also tackled
how to better shape the workforce for the new trends and how we should approach education and public outreach.
This document was coauthored during the workshop and edited in the weeks after.
It comprises the discussions and highlights many recommendations which came out of the workshop.

 We shall distill parts of this document and  formulate potential white papers for the decadal survey.
\end{abstract}

\maketitle

\tableofcontents

\section{Introduction\Contact{William O'Mullane}}\label{sec:intro}

\Contributors{William O'Mullane <\mail{womullan@lsst.org}>, Ting Li <\mail{tingli@fnal.gov}>}
\bigskip


In the Petabyte era the lines between \gls{software}, technology and science are blurred - the chance to do science with petabytes without major infrastructure is pretty slim. Therefore the importance of technology in science exploitation becomes ever more important, which also implies we pick up the pace in training the workforce and in the areas of education and public outreach.

The Kavli foundation sponsored a series of workshops on the theme \emph{Petabytes to Science}\footnote{\url{https://petabytestoscience.github.io/}}, the second of which was held 12$^{th}$ to 14$^{th}$ of February 2019 in Las Vegas.
The aim of the this second workshop was to formulate potential \gls{APC} white papers. To facilitate this we
 discussed important trends,  technologies, approaches to workforce management, education and public outreach.
We took a holistic approach and built a single document encompassing several broad categories, namely:
\begin{itemize}
	\item Science drivers (\secref{sec:science}) - which science cases need new techniques and approaches?
	\item Data Management (\secref{sec:data}) - what data management challenges does this present?
	\item Software (\secref{sec:software}) - how should \gls{software} be developed to meet those challenges?
	\item Technology and Infrastructure (\secref{sec:tech}) - what technologies and infrastructure is needed to under pin the services?
	\item Workforce and Inclusion (\secref{sec:iwft}) - what training should we do to prepare ? How can we improve and diversify the workforce?
	\item Education and Public Outreach (\secref{sec:epo}) - through \gls{EPO} can we increase awareness of the public about astronomy and ensure future finding streams? What are the challenges and opportunities for EPO?

\end{itemize}

From each of the sections a number of recommendations were identified, these are summarized in \secref{sec:recs}.
For each recommendation we suggest the audiences they are useful to and the time period in which they should be executed (this may be seen as a sort of priority). The time periods or terms are short term (1-3 years), medium term (3-5 years) and long term (5-10 years).

The intention is to extract some decadal survey \gls{APC} papers from these ideas.
If you are interested in contributing to or endorsing
\href{https://tinyurl.com/y2ksemp2}{white papers on these topics sign up here}\footnote{\url{https://tinyurl.com/y2ksemp2}}  or contact the authors listed in this document -- names followed by an email address are the leads of each chapter.

\noindent\textbf{Note:} This document is a collection of ideas and a record from a workshop  - it is not a polished document.  We
have made some effort to not have repetitions between chapters however we do not guarantee a pleasant coherent read.

\section{Recommendations} \label{sec:recs}


These tables summarize the recommendations in the document per audience we feel would be interested. Clicking the label or text will take you to the full recommendation in the document.\footnote{In a \gls{PDF} it may be useful to note that CMD $\leftarrow$ (CTRL on Windows/Linux) returns you to from whence you clicked.}
Please note that a recommendation may be aimed at multiple audiences and therefore may appear more than once in the tables below.
\tiny \begin{longtable} { p{0.78\textwidth}  p{0.13\textwidth}  p{0.09\textwidth} } 
\caption{Astronomer recommendations. \label{tab:recAstronomer1}}\\ 
\hline 
\textbf{Recommendation }&\textbf{Area }&\textbf{Term } \\ \hline
\hyperref[rec:datastand]{{\textbf \tiny REC-1}  Adopt common data models throughout the astronomical community }&Data Management &Short  \\ 
\hyperref[rec:dataprop]{{\textbf \tiny REC-3}  Proprietary data time scales should be limited, and all datasets should be eventually made publicly available }&Data Management &Short  \\ 
\hyperref[rec:opspractices]{{\textbf \tiny REC-8}  Improve long-term \gls {software} and service support }&Technology &Short  \\ 
\hyperref[rec:swmoney2]{{\textbf \tiny REC-11}  Funding for sustaining core astronomical ``community infrastructure'' projects }&Software &Medium  \\ 
\hyperref[rec:swecoplan]{{\textbf \tiny REC-12}  Cultivating a sustainable research \gls {software} ecosystem }&Software &Short  \\ 
\hyperref[rec:algmodels]{{\textbf \tiny REC-13}  Create funding models and programs to support the development of advanced algorithms and statistical methods specifically targeted to the astronomy domain }&Analysis &Medium  \\ 
\hyperref[rec:discengine]{{\textbf \tiny REC-14}  Build automated discovery engines }&Analysis &Long  \\ 
\hyperref[rec:algconnect]{{\textbf \tiny REC-15}  Promote interdisciplinary collaboration between institutions, fields, and industry }&Analysis &Long  \\ 
\hyperref[rec:algeducation]{{\textbf \tiny REC-16}  Develop an open educational curriculum and principles for workforce training in both algorithms and statistics }&Analysis &Medium  \\ 
\hyperref[rec:algpub]{{\textbf \tiny REC-17}  Encourage, support, and require open publication and distribution of algorithms }&Analysis &Short  \\ 
\hyperref[rec:wc6]{{\textbf \tiny REC-22}  Software training as part of science curriculum }&Workforce &Medium  \\ 
 \hline
\end{longtable} \normalsize

\tiny \begin{longtable} { p{0.78\textwidth}  p{0.13\textwidth}  p{0.09\textwidth} } 
\caption{Manager recommendations. \label{tab:recManager1}}\\ 
\hline 
\textbf{Recommendation }&\textbf{Area }&\textbf{Term } \\ \hline
\hyperref[rec:datapreserve]{{\textbf \tiny REC-4}  Long-term data preservation of datasets }&Data Management &Long  \\ 
\hyperref[rec:swecoplan]{{\textbf \tiny REC-12}  Cultivating a sustainable research \gls {software} ecosystem }&Software &Short  \\ 
\hyperref[rec:algconnect]{{\textbf \tiny REC-15}  Promote interdisciplinary collaboration between institutions, fields, and industry }&Analysis &Long  \\ 
\hyperref[rec:wcA]{{\textbf \tiny REC-25}  Recognize \gls {software} as part of the career path }&Workforce &Short  \\ 
 \hline
\end{longtable} \normalsize

\tiny \begin{longtable} { p{0.78\textwidth}  p{0.13\textwidth}  p{0.09\textwidth} } 
\caption{University recommendations. \label{tab:recUniversity1}}\\ 
\hline 
\textbf{Recommendation }&\textbf{Area }&\textbf{Term } \\ \hline
\hyperref[rec:wc6]{{\textbf \tiny REC-22}  Software training as part of science curriculum }&Workforce &Medium  \\ 
\hyperref[rec:wcB]{{\textbf \tiny REC-26}  Partnerships to support data science staff }&Workforce &Medium  \\ 
 \hline
\end{longtable} \normalsize

\tiny \begin{longtable} { p{0.78\textwidth}  p{0.13\textwidth}  p{0.09\textwidth} } 
\caption{Agency recommendations. \label{tab:recAgency1}}\\ 
\hline 
\textbf{Recommendation }&\textbf{Area }&\textbf{Term } \\ \hline
\hyperref[rec:datalowbar]{{\textbf \tiny REC-2}  Eliminate barriers to public data access }&Data Management &Medium  \\ 
\hyperref[rec:dataprop]{{\textbf \tiny REC-3}  Proprietary data time scales should be limited, and all datasets should be eventually made publicly available }&Data Management &Short  \\ 
\hyperref[rec:datapreserve]{{\textbf \tiny REC-4}  Long-term data preservation of datasets }&Data Management &Long  \\ 
\hyperref[rec:cbra]{{\textbf \tiny REC-5}  Develop a community wide architecture supporting Science as a Service }&Technology &Long  \\ 
\hyperref[rec:fundops]{{\textbf \tiny REC-7}  Enable support for full mission life cycle including long-term data products  }&Technology &Short  \\ 
\hyperref[rec:opspractices]{{\textbf \tiny REC-8}  Improve long-term \gls {software} and service support }&Technology &Short  \\ 
\hyperref[rec:cmdeploy]{{\textbf \tiny REC-9}  Fund cross-mission deployment }&Technology &Medium  \\ 
\hyperref[rec:swmoney1]{{\textbf \tiny REC-10}  Funding for \gls{software} development in existing grant programs }&Software &Long  \\ 
\hyperref[rec:swmoney2]{{\textbf \tiny REC-11}  Funding for sustaining core astronomical ``community infrastructure'' projects }&Software &Medium  \\ 
\hyperref[rec:swecoplan]{{\textbf \tiny REC-12}  Cultivating a sustainable research \gls {software} ecosystem }&Software &Short  \\ 
\hyperref[rec:algmodels]{{\textbf \tiny REC-13}  Create funding models and programs to support the development of advanced algorithms and statistical methods specifically targeted to the astronomy domain }&Analysis &Medium  \\ 
\hyperref[rec:algconnect]{{\textbf \tiny REC-15}  Promote interdisciplinary collaboration between institutions, fields, and industry }&Analysis &Long  \\ 
\hyperref[rec:algeducation]{{\textbf \tiny REC-16}  Develop an open educational curriculum and principles for workforce training in both algorithms and statistics }&Analysis &Medium  \\ 
\hyperref[rec:algpub]{{\textbf \tiny REC-17}  Encourage, support, and require open publication and distribution of algorithms }&Analysis &Short  \\ 
\hyperref[rec:wc3]{{\textbf \tiny REC-18}  Programs to cultivate the next generation }&Workforce &Long  \\ 
\hyperref[rec:wc5]{{\textbf \tiny REC-20}  Long-term curation of materials }&Workforce &Long  \\ 
\hyperref[rec:wcfundpartner]{{\textbf \tiny REC-21}  Funding for innovative partnerships }&Workforce &Medium  \\ 
\hyperref[rec:wc1]{{\textbf \tiny REC-23}  Training activities and materials }&Workforce &Short  \\ 
\hyperref[rec:wcFund]{{\textbf \tiny REC-27}  Support long-term technical capacity }&Workforce &Medium  \\ 
 \hline
\end{longtable} \normalsize

\tiny \begin{longtable} { p{0.78\textwidth}  p{0.13\textwidth}  p{0.09\textwidth} } 
\caption{Educator recommendations. \label{tab:recEducator1}}\\ 
\hline 
\textbf{Recommendation }&\textbf{Area }&\textbf{Term } \\ \hline
\hyperref[rec:wc6]{{\textbf \tiny REC-22}  Software training as part of science curriculum }&Workforce &Medium  \\ 
 \hline
\end{longtable} \normalsize

\tiny \begin{longtable} { p{0.78\textwidth}  p{0.13\textwidth}  p{0.09\textwidth} } 
\caption{Technologist recommendations. \label{tab:recTechnologist1}}\\ 
\hline 
\textbf{Recommendation }&\textbf{Area }&\textbf{Term } \\ \hline
\hyperref[rec:datastand]{{\textbf \tiny REC-1}  Adopt common data models throughout the astronomical community }&Data Management &Short  \\ 
\hyperref[rec:cbra]{{\textbf \tiny REC-5}  Develop a community wide architecture supporting Science as a Service }&Technology &Long  \\ 
\hyperref[rec:coloc]{{\textbf \tiny REC-6}  Enable new scales of research through data co-location }&Technology &Medium  \\ 
\hyperref[rec:opspractices]{{\textbf \tiny REC-8}  Improve long-term \gls {software} and service support }&Technology &Short  \\ 
\hyperref[rec:swmoney2]{{\textbf \tiny REC-11}  Funding for sustaining core astronomical ``community infrastructure'' projects }&Software &Medium  \\ 
\hyperref[rec:discengine]{{\textbf \tiny REC-14}  Build automated discovery engines }&Analysis &Long  \\ 
 \hline
\end{longtable} \normalsize





\section{Scientific Context and Drivers}\label{sec:science}


\noindent\Contributors{Adam Bolton <\mail{bolton@noao.edu}>, Eric Bellm, Alex Drlica-Wagner, Ting Li, Raffaella Margutti, Gautham Narayan, Meg Schwamb}
\bigskip

\noindent\textbf{Note:}
If you have come directly to this chapter we suggest you please read at least the Introduction in \secref{sec:intro} before delving further.
\newline

The last two decades have seen a significant increase in the prominence of data-intensive, survey-scale astronomy. Surveys such as \gls{SDSS}, \gls{DES}, \gls{Pan-STARRS}, and \gls{ZTF} have pioneered these modes. Even more ambitious projects such as \gls{DESI}, \gls{LSST}, \gls{WFIRST}, and \gls{SKA} are rapidly approaching, bringing new opportunities and challenges for petascale astronomical science.

From an experimental design perspective, the development of large survey projects and facilities has been driven by fundamental scientific questions about the Solar System, our Milky Way and its stellar populations and satellites, the evolution of galaxies and quasars, and the nature of dark matter and dark energy. These questions have in common the need to obtain significant statistics over large population samples or volumes, or alternatively, to realize significant probabilities for the discovery of rare objects or events.

Big surveys naturally lead to big datasets. These big datasets in turn bring qualitatively new challenges in data management, computing, \gls{software}, and professional development that must be tackled to realize the scientific promise of the surveys themselves.

Big datasets from big surveys also open up diverse opportunities for \textit{data-driven science}: research and discovery programs defined entirely on the basis of available datasets, not on the basis of collecting new data. This approach is especially empowering of exploratory science, as described in the series of Astro2020 white papers by \citet{Fabbiano_wp_103,Fabbiano_wp_102,Fabbiano_wp_101,Fabbiano_wp_100,Fabbiano_wp_98,Fabbiano_wp_99}. Data-driven research can multiply the scientific impact of a survey, and can be especially effective for broadening participation in forefront astronomical research beyond those groups with the greatest access to resources. Data-driven science with large surveys calls on many of the same data-intensive methods as are required for ``primary'' survey science, while also presenting new challenges and requirements such as public data release and broad data accessibility.

In the following subsections, we outline some  current scientific opportunities, and their associated data-intensive challenges, across a broad range of astrophysics and cosmology drawn from science white papers submitted to the Astro2020 Decadal Survey. In the subsequent chapters of this report, we address the crosscutting technology, methods, and professional considerations that will support success in these scientific areas in the next decade.

\subsection{Planetary Systems; Star and Planet Formation}

LSST will conduct a 10-year survey across the southern sky, revisiting the same locations approximately every three days. This time-resolved dataset will detect both transient objects in the fixed sky and moving objects in the Solar System. \citet{Chanover_wp_378} describe the promise of \gls{LSST} for the discovery of \textit{dynamic} Solar System phenomena such as active asteroids and small-body collisions. To yield their scientific potential, these objects require rapid detection, alert, and follow-up observation. This implies the need for a coordinated real-time \gls{software} infrastructure beyond the scope of \gls{LSST} operations deliverables. Similarly, \citet{Holler_wp_79} describe the Solar System science potential of \gls{WFIRST}, which will require the deployment of robust moving-object detection algorithms within the \gls{WFIRST} data management framework.

A core goal of the \gls{WFIRST} mission is to conduct a microlensing census of extrasolar planets. \citet{Yee_wp_425} and \citet{Gaudi_wp_235} describe both core and ancillary science potential of this aspect of \gls{WFIRST}, which highlights the algorithmic and software-systems engineering challenge of addressing diverse microlensing applications within a petascale dataset with quality comparable to space-based telescopes.

\citet{Ford_wp_517} discuss the essential role of advanced statistical and machine-learning methodologies for optimal extraction of Doppler signatures of extrasolar planets with high-resolution spectroscopy in the coming decade. As the experimental forefront approaches the 10 cm\,s$^{-1}$ precision necessary to detect true Earth analogs around Sun-like stars, new statistics and algorithms become ever more crucial.

\subsection{Stars and Stellar Evolution; Resolved Stellar Populations}

\citet{Pevtsov_wp_212} highlight the potential scientific return for stellar astrophysics from digitizing historical astronomy data and making it available in accessible forms within modern data-management systems.

\citet{Dey_wp_541} and \citet{Kollmeier_wp_557} describe the potential for data-mining within large spectroscopic survey datasets (e.g. \gls{SDSS}, \gls{DESI}) to discover primordial Population III stars as well as new, rare, and unexpected stellar types.

Several Astro2020 white papers highlight the scientific potential that arises from combining multiple large-scale resolved stellar datasets. Asteroseismology results can be sharpened through the combination of time-series data from \gls{TESS}, \gls{PLATO}, and \gls{WFIRST} with stellar spectroscopic parameters measured by SDSS-V, \gls{MSE}, and other surveys \citep{Huber_wp_540}. Our knowledge of the structure, formation, stellar populations, and cosmological context of the Milky Way will be maximized through the combination of photometry, astrometry, and spectroscopy from multiple survey missions \citep{Sanderson_wp_387,Williams_wp_337}. Joint time-resolved analysis of photometric, astrometric, and spectroscopic survey data will also enable diverse astrophysical applications of stellar multiplicity \citep{Rix_wp_119}. For all these scientific goals to be realized, full \gls{interoperability} and combined analysis at the scale of millions to billions of stars will be required across all relevant surveys, which poses significant challenges in data management and \gls{software} systems, as described by \citet{Olsen_wp_372}.

\subsection{Compact Objects; Time-Domain and Multi-Messenger Astrophysics}

\citet{Graham_wp_377} highlight the explosive-transient discovery-space potential for \gls{LSST} combined with \gls{ELT} follow-up, which will only be realized if detection, filtering, and follow-up can be triggered rapidly by scientifically tuned \gls{software} systems. \citet{Kirkpatrick_wp_123} argue for increasing the transient and variable science return of the NEOCam mission through investment in the \gls{software} infrastructure needed to detect, monitor, and alert on non-moving (i.e. non-Solar system) variable sources. This is an example of how \gls{software} alone can qualitatively change the scientific opportunity space of a given survey/mission.

\citet{Cowperthwaite_wp_402} argue for the importance of ``target-of-opportunity'' observing with \gls{LSST} to follow up on \gls{LIGO} gravitational wave triggers in search of a counterpart. This points to the need for sophisticated real-time data-management \gls{software} systems such as \gls{LSST}'s to be implemented in ways that are flexible to the development of new and potentially unanticipated operational modes.

Binary systems with compact-object components provide ``astrophysical laboratories'' that will be central to many time-domain and multi-messenger applications in the coming decade. \citet{Maccarone_wp_251} describe the scientific potential for increasing the sample of known stellar binaries with a black hole component, and highlight the importance of time-domain photometric surveys such as \gls{LSST}, \gls{ZTF}, \gls{ATLAS}, PanSTARRS for identifying candidate systems through analysis of light curves to identify ellipsoidally modulated binaries and optically-outbursting X-ray binaries. \citet{Eracleous_wp_16} describe the role of \gls{LSST} and \gls{ZTF} in catching the disruption of white dwarf stars by a black hole companion, which can be further informed by \gls{LISA} observations when available. \citet{Littenberg_wp_44} and \citet{Kupfer_wp_210} describe the importance of \gls{LSST}, \gls{ZTF}, Gaia, BlackGEM, SDSS-V, and \gls{DESI} for identifying ultracompact binaries that will be potential future persistent gravitational wave sources for \gls{LISA}. In all these cases, algorithmic time-domain analysis and discovery implemented through science-driven \gls{software} systems will be essential.

\citet{Palmese_wp_346} highlight the potential for large spectroscopic surveys to provide redshifts for hosts of future gravitational-wave inspiral sources, both ``bright'' and ``dark''. These redshifts will enable ``standard siren'' cosmology in combination with the inferred intrinsic parameters of the \gls{GW} sources. This points to the need for spectroscopic surveys to make their data archives fully accessible.

Cutting across several of the scientific topics above, \citet{Chang_wp_482} provide an overview of ``cyberinfrastructure'' needs for multi-messenger astrophysics in the coming decade.

\subsection{Galaxy Evolution}

Large extragalactic surveys and their associated data archives are a key resource for advancing our understanding of galaxy evolution. \citet{Behroozi_wp_141} highlight the importance of large surveys, accessible data archives, and open \gls{software} for advancing our knowledge of galaxy evolution through the particular method of empirical modeling. \citet{Dickinson_wp_595} envision a future spectroscopic galaxy survey that would provide a highly complete SDSS-like sample across multiple redshifts, which would enable a comprehensive study of the coevolution of galaxies and their stellar populations with the formation of dark matter halos across cosmic time. Additional galaxy-evolution (and cosmology) science drivers are discussed below in Section~\secref{sec:multiprobe} in the context of combining data from multiple surveys and facilities.

Multiple quasar science opportunities in the next decade will be driven by survey-scale and data-intensive methodologies. \citet{Shen_wp_302} describe the role of large, time-resolved spectroscopic surveys to map the structure and growth of quasars through the method of reverberation mapping. \citet{Fan_wp_137} highlight the prospects for data mining in \gls{LSST} and \gls{WFIRST} for large samples of high-redshift luminous quasars, which can probe the coevolution of galaxies and their central SMBHs at early times. \citet{Pooley_wp_455} highlights the role of \gls{LSST} as a resource for discovery of strongly lensed quasars which can uniquely probe the dark-matter fraction in the lensing galaxy, while \citet{Moustakas_wp_539} describe the role that these same systems can play in reconstructing the detailed structure of quasars themselves.

\citet{Lehner_wp_524} highlight the importance of high-quality spectroscopic reduction pipelines and accessible data archives to maximize the science potential of high-resolution spectroscopy on large ground-based telescopes to trace the evolution of the intergalactic and circumgalactic medium over cosmic time.

\subsection{Cosmology and Fundamental Physics}

With its goal of understanding the contents and evolution of the universe as a whole, cosmology has driven many of the recent advances in ``big data'' astronomy. This trend is likely to continue through the 2020s. Many Astro2020 science white papers describe planned and proposed missions for which robust data-processing and data-management systems will be essential baseline requirements. These projects require not just basic management of petascale data, but also the automated execution of sophisticated inference algorithms---for galaxy shapes, photometric and spectroscopic redshifts, selection functions---across their entire datasets.

\citet{Slosar_wp_111} provide a broad overview of prospects for ongoing study of dark energy and cosmology with large-scale surveys. \citet{Dore_wp_379} give an overview of the cosmological science capabilities of \gls{WFIRST} via the channels of weak lensing, galaxy clustering, supernovae, and redshift-space distortions. \citet{Wang_wp_563} describe the dark-energy science potential of multi-tracer wide-field spectroscopic surveys that achieve higher completeness and spatial density than existing or planned surveys. \citet{Slosar_wp_112}, \citet{Ferraro_wp_85}, and \citet{Meeburg_wp_122} describe the prospects for constraining models of inflation and early-Universe physics through the signatures of primordial non-Gaussianity in large-scale structure surveys. \citet{Pisani_wp_51} describe the potential to constrain dark energy, neutrinos, and modified gravity in cosmic voids within densely sampled redshift surveys. \citet{Dvorkin_wp_75} describe the prospect for measuring the absolute neutrino mass scale through several large-scale observational channels. \citet{Rhodes_wp_129} envision the definitive large-scale structure survey to map the three-dimensional position of all galaxies and dark-matter halos in the visible universe. \citet{Geach_wp_608} envision a future wide-field spectroscopic survey in the sub-millimeter which would cover redshifts 1--10.

Large surveys and their associated \gls{software} and data systems will likewise be central to the quest to understand dark matter in the coming decade. \citet{Gluscevic_wp_150} describe the potential for galaxy and Lyman-alpha forest surveys, in combination with modeling and simulation of baryonic effects, to constrain the nature of particle dark matter. \citet{Bechtol_wp_230} describe the dark-matter science potential of \gls{LSST} on its own and in combination with spectroscopic facilities. Other channels for constraining particle dark matter with large spectroscopic surveys of galaxies and Milky Way stars are described by \citet{Li_wp_279}. \citet{grin_wp_XXX} describe how a combination of \gls{CMB}, optical, infrared, and gravitational wave observations will contribute to our understanding of ultra-light dark matter candidates.

Survey-scale cosmological science in the 2020s will also leverage machine learning (\gls{ML}) supported by large and well-calibrated datasets. \citet{Ntampaka_wp_21} describe recent applications of \gls{ML} to a diverse range of applications in cosmology, and highlights some of the most significant opportunities for \gls{ML} to increase the scientific return from \gls{LSST}, \gls{SKA}, and other major future projects.

\subsection{Combining Multiple Probes of Cosmology and Galaxy Evolution}
\label{sec:multiprobe}

A central theme of many Astro2020 science white papers at the interface of galaxy evolution and cosmology---and one that will significantly drive requirements for the computing, data, and \gls{software} systems of the 2020s---is the need to combine and co-analyze data from multiple major surveys. These use cases imply requirements for data accessibility, \gls{interoperability}, and mobility between data-hosting locations. They will also drive the astronomy and cosmology communities to leverage the capabilities of research-supercomputing and commercial-cloud computing providers in new ways.

\citet{Newman_wp_399} and \citet{Mandelbaum_wp_404} describe the synergistic potential for deep and wide-field survey spectroscopy to enhance the dark-energy science return from \gls{LSST}. \citet{Chary_wp_55}, \citet{Eifler_wp_463}, and \citet{Rhodes_wp_224} describe joint analysis approaches for \gls{LSST}, Euclid, and \gls{WFIRST} that would enhance the resulting weak-lensing and galaxy-clustering cosmology measurements of these missions. \citet{Capak_wp_521} describe the scientific benefit from coordination of ``deep field'' regions across multiple surveys and multiple wavelengths. (Here, standardized data and metadata formats will be necessary not only to realize the scientific potential of diverse datasets in common areas of sky, but also to enable discovery of existing datasets and coordination of planned future deep-field campaigns.) \citet{Furlanetto_wp_161}, \citet{Cooray_wp_59}, and \citet{Cuby_wp_401} describe the potential for combining galaxy surveys (space and ground-based), 21cm surveys, and other probes to obtain a more detailed picture of the epoch of reionization. \citet{Mantz_wp_308} describe the importance of combining multiple large surveys across wavelength for the selection of uniform and significant samples of high-redshift galaxy clusters.

\section{Data Management}\label{sec:data}
\noindent\Contributors{Anne-Marie Weijmans <\mail{amw23@st-andrews.ac.uk}>, JJ Kavelaars <\mail{JJ.Kavelaars@nrc-cnrc.gc.ca}>, Surajit Chaudhuri, Vandana Desai, Jamie Kinney, William O'Mullane, Alex Szalay}

\bigskip
\noindent\textbf{Note:}
If you have come directly to this chapter we suggest you please read at least the Introduction in \secref{sec:intro} before delving further.
\newline

In this section we recognize two of the main challenges related to data management in the next decade:

\begin{itemize}
    \item Big Data: datasets will be of such large volume, that moving them across individual data repositories is not practical. This will affect the way that we interact with data, and has the risk that some users will be excluded from access to large datasets. (See also \secref{sec:coloc})
    \item Time Domain: datasets will contain a time domain element, i.e. will contain data of the same part of the sky obtained at different time intervals. This will put challenges on current visualisation and discovery tools.
\end{itemize}

To address these two challenges, we make the following recommendations:

    \nrec{Data Management}{Astronomer, Technologist}{Short}{datastand}{Adopt common data models throughout the astronomical community}{The astronomical community should work towards a common data model. This will allow astronomers to concentrate on scientific exploration of datasets, without having to worry about data formats and structures}
    \nrec{Data Management}{Agency}{Medium}{datalowbar}{Eliminate barriers to public data access}{Astronomical public datasets should be accessible to everyone, and everyone should have the opportunity to contribute to astronomical public datasets}
   \nrec{Data Management}{Agency, Astronomer}{Short}{dataprop}{Proprietary data time scales should be limited, and all datasets should be eventually made publicly available}{To maximize scientific output, and allow wider-community access of centralized funded projects, all astronomical datasets should be made publicly available and accessible after an appropriate but short proprietary time limit}
    \nrec{Data Management}{Agency, Manager}{Long}{datapreserve}{Long-term data preservation of datasets}{Long-term data preservation and management should be an integral part of community-wide project planning}


We discuss these recommendations in more detail in the sections below.

\subsection{Interoperability}


In the 2020s, new datasets such as \gls{LSST}, \gls{WFIRST}, and Euclid hold enormous science promise. Realizing this potential for transformative science presents a number of challenges for data management. As we outlined above: the main challenges are the volumes of the data, as well as the additional dimension that time domain observations will bring to these large datasets.

\subsubsection{Common observation models}

{\em Data centers should adopt a common observation model (\recref{rec:datastand}}). A common observation model is a set of standard \gls{metadata} parameters that can be used to describe any astronomical dataset. The widespread adoption of a common observation model has many advantages, outlined below.

The large volume of data in the 2020s implies that re-processing these data will incur high cost, thus increasing sharply the importance of a common data model that can serve as the basis for information exchange and reuse. International Virtual Observatory Alliance \gls{IVOA} is on the way to adopting the Common \gls{Archive} Observation Model (\gls{CAOM}) for images. It has already been adopted by a number of large archives, including the \gls{CADC}, the \gls{ESAC}, the Mikulski \gls{Archive} for Space Telescopes (\gls{MAST}), and the \gls{NASA}/\gls{IPAC} Infrared Science \gls{Archive} (\gls{IRSA}).

Effective re-use of data requires careful, ongoing curation of this \gls{metadata} model. This includes both preserving the expertise and context of what the nuances of a particular dataset are, but also periodically updating \gls{metadata} to conform to new standards and meet new use cases. For example, \gls{astrometry} of old datasets may need to be updated to support real-time querying/matching/aligning/jointly processing for time domain studies.

\subsubsection{Data storage}

{\em Data Centers should adopt industry standards for data storage when possible,} see also \secref{sec:coloc}.
Perhaps the most obvious challenge is simply storing the data. The large volumes mean efficiency in storage representation is important.

We recommend that data centers leverage ‘off-the-shelf’, open data management services, tools, and technologies that have been developed by industry.
Moving to industry standards for things like images allow us to leverage new technologies such as the ability to stream and operate remotely on {\em objects} using standard tools.
File systems as we know them will not be the most appropriate storage model at petascale levels.
Alternatives include the use of \gls{cloud} object stores, \gls{cloud} compute, ‘big data native’ formats such as Apache \gls{parquet} and \gls{OpenEXR}, and cloud-optimized \gls{FITS} (see \gls{cloud} optimized GeoTIFF as an example https://www.cogeo.org). Traditional astronomy file formats (e.g. \gls{FITS}) should be used as they were originally intended, for transport only. That being said, one big advantage of \gls{FITS} files is their ability to co-package meta-data, while e.g. for Parquet there are only limited options to have meta-data included directly with the data. Data and meta-data should be managed together to not lose efficiency in analysis performance.

\subsubsection{Eliminating file systems}

{\em The community should develop a flexible suite of application program interfaces to abstract the file system.}

The previous recommendation calls for using storage formats that are optimized for the \gls{cloud}, in order to meet the challenge of “Big Data” storage. This also implies that the current often used practice of storing files and file systems locally on astronomers' laptops of analysis will have to change to this more global approach of accessing and analyzing data remotely. To avoid a difficult transition for many individual astronomers, global file structures and formats should be abstracted by two layers of application program interfaces (\gls{API}s). The bottom layer consists of a limited set of (\gls{VO})-based APIs implemented by data centers. We recommend that data centers implement a critical set of core \gls{VO} \gls{API}s, including cone search, image search, spectral search, \gls{TAP}, the standardized language used to report observations of astronomical events \gls{VO}Event, and Ephemeris Lookup (still to be adopted by the \gls{IVOA}). Other \gls{VO} standard protocols have become obsolete, and should not be implemented (e.g. \gls{VO}Space in favor of \gls{S3}). The top layer consists of user-facing \gls{API}s developed by the community to “hide” the file formats from the user. In the 2020s, this top layer should focus on Python. However, lightweight \gls{API}s can be built in other languages as community needs dictate.


\subsubsection{Interoperable science platforms}

{\em Data Centers should provide a set of interoperable science platforms.}

A science platform provides users with access to compute and analytic services close to the datasets of interest. With new astronomy survey datasets measured in petabytes, it is quickly becoming infeasible to copy entire datasets to another location for analysis. At the same time, it is increasingly common for researchers to leverage big data and inefficient parallel compute technologies to analyze large subsets, if not entire datasets. Cloud services provided by commercial organizations and \gls{DOE}/\gls{NSF}-funded High Performance Computing (\gls{HPC}) centers offer both the scale of compute resources and the networking infrastructure required to analyze these datasets using modern techniques.  Furthermore, by physically co-locating datasets, we make it possible for researchers to conduct investigations that incorporate data from multiple mission archives. Therefore, we recommend that the \gls{DOE}, \gls{NSF}, and other funding agencies encourage data archives to be physically stored and perhaps co-located in facilities which are accessible to the global research community and which provide the compute and higher-level analytical services that will be used to analyze these datasets at scale.

\subsection{Lowering the barriers to public data access}

{\em Projects should eliminate barriers to public data access (\recref{rec:datalowbar}), and limit the proprietary data time scale (\recref{rec:dataprop}).}
To make maximal use of astronomical datasets, every astronomer should have access to these datasets, and have the tools available to exploit their scientific richness. In the sections below we make suggestions that projects should adopt to ensure that barriers to work with data are removed: we concentrate here on astronomical community (including students), and refer to \secref{sec:epo} for promoting astronomical data with the general public. We also recommend that although proprietary data has its use within the astronomical community (e.g. ensuring that the astronomers and students who invested in collecting and reducing the data have the opportunity to explore the datasets for their science), that these proprietary times are kept short to maximize over-all science output.

\subsubsection{Computational resources}
{\em Projects should make their datasets available for remote data analysis}
As mentioned in the previous section, the big datasets of the next astronomical surveys will be too large to download and store on individual astronomers computing systems (laptops). Projects should therefore ensure that their data is available for remote analysis, and provide opportunities for \gls{cloud} computing. This will ensure that the whole astronomical community will have access to the data, and that lack of large data storage and/or computing facilities will not prevent astronomers from taking part in the scientific exploration of large datasets. We note that \gls{cloud} computing does require reliable internet connections, which for most of the astronomical community will be available, but not necessarily for a more general audience (e.g. schools and individuals in remote areas).

\subsubsection{Documentation}
{\em Projects should allocate sufficient resources and attention to capturing the expertise on collection, processing and interpretation of their data products.}
A dataset is only as strong as its documentation. Without documenting the expertise needed to work with a dataset, scientific analysis based on that data has a high risk of being flawed. Including detailed documentation with data resources is therefore a must.  The documentation that captures the projects expertise should be easily accessible: the documentation should be released at the same time as the datasets. The documentation should be clearly written, with jargon explained and with tutorials and examples for clarification. The documentation should not be aimed at the experts within a project, but be written with inexperienced, new users in mind (e.g. students). There should be a mechanism (e.g. helpdesk, forum), in place to collect feedback and errata, and the documentation should be updated and improved accordingly during the life time of the project. Having excellent documentation does not only lower the barriers of entry to work with large datasets, but will also be invaluable when the project has reached the end of its lifetime, and the datasets will (eventually) go into long-term archiving (see Section~\secref{sec:datapreservation}).

\subsubsection{Professional training}
{\em Training resources on the exploration of large public datasets should be made available for free and on-line.}
To lower barriers for entry further, there should be training resources available for the astronomical community, to ensure that they can explore the richness of large public datasets. These training resources, such as tutorials, demos and notebooks, should be aimed at appropriate levels, as the education needs of a beginning students are different than those of a postdoc or faculty member. These resources should be available and accessible to a large audience, and therefore should be linked to from dataset documentation pages.



\subsubsection{Education and Public Outreach}
{\em Data facilities should invest in collaborations with Education and Public Outreach teams.}
Having real astronomical public data available for education and public outreach purposes is a big advantage for developing resources that closely mimic and can even contribute to scientific research.
As outlined in  \secref{sec:epo} the Education and Public Outreach (\gls{EPO}) chapter of this document, we recommend supporting dedicated education and outreach groups with relevant expertise to maximize the impact of \gls{EPO} activities. To work closely with these \gls{EPO} teams, we recommend that  each data facility has at least one team member to liaise with the \gls{EPO} team, and provide input on data requirements for \gls{EPO} activities.

\subsection{Long-term data preservation}
\label{sec:datapreservation} 

{\em Long-term data preservation and management should be an integral part of community-wide project planning (\recref{rec:datapreserve})}
Data that is actively used will continue to exist in the community: there is a sort of Darwinian selection going on constantly. Expertise is therefore also kept reasonably current while the data are in use. But the implication is that some data will be getting used less and less over time, and at some point is going to be compressed (including documentation and possible email archives) and put into \gls{cold storage} for long term preservation. Catalogs and derived data products could potentially persist in regular use for longer than their source data.

The long term preservation of data is a problem that is not unique to Astronomy or Science, neither in volume nor in characteristics. Such preservation has the following components:
\begin{enumerate}
    \item Ensuring data integrity (no tampering)
    \item Sufficient redundancy so that there is no single point of failure, to ensure data access
    \item Packaging of information that provides “recoverability” of essential information
    \item Funding for such preservation, as well as data format and \gls{software} maintenance.
\end{enumerate}
The first challenge (data integrity) is  a general problem, and  there are many techniques that have been developed in research and in industry to ensure integrity. These include tamper-proof logs and signature based comparison of multiple copies of preserved data including watermarking. We should select a preferred method in astronomy.

The second challenge is perhaps met by having multiple sites to ensure that there is no single point of failure.  There is a cost vs. “how many failures you can tolerate” trade-off. Offloading this task to multiple vendors of public \gls{cloud} companies is probably the simplest solution. One compelling reason to do that is because they will, due to market pressure, continue to support changing data formats and media as technology change. Through all these, the data should remain accessible, even when in \gls{cold storage}.

The third challenge, which is packaging of information, is most critical and this is where unique aspects of Astronomy are relevant. A data dump in itself is not easy to interpret, especially after several years, when the experts that generated and worked with the data have moved on to other projects. Therefore, it is critical to have good documentation and \gls{metadata} enrichment. We need to  capture the \emph{expertise} so we may want to compress and store communications such as Slack channels, mailing lists and logs etc. As well as the actual data.
By having such additional catalogs, the “recoverability” of value from preserved data is much enhanced. However, we need to acknowledge that such often more informal and unsorted communication does not replace the need for comprehensive and understandable documentation and tutorials to work with the data. The value of archived communication would be for the (hopefully) rare instances that an issue occurs that is not documented probably, but was discussed on communication channels, and for historic and/or social studies.

Last but not the least, there is the funding question. There are two possible models. First model is to attach a “service fee” to every funded project to support ongoing high quality documentation. Alternatively, funding may be requested as we near end of the project. The payment model, especially to \gls{cloud} providers, could be fashioned like what is done for title insurance for home purchases – an one-time payment for a fixed number of years

\section{Technology \& Infrastructure \Contact{William O'Mullane, Niall Gaffney}}\label{sec:tech}

\Contributors{William O'Mullane <\mail{womullan@lsst.org}>, Niall Gaffney <\mail{ngaffney@tacc.utexas.edu}>, JJ Kavelaars, Frossie Economou, Surajit Chaudhuri}

\bigskip
\noindent\textbf{Note:}
If you have come directly to this chapter we suggest you please read at least the Introduction in \secref{sec:intro} before delving further.
\newline

We discussed many of the challenges and potential technological and infrastructure innovations which could be future
solutions to current problems.
In these discussions, we concluded that the goal was not to predict future
problems and unknown technological solutions, but to unite and align
cross mission and research community \gls{cyberinfrastructure} needs.
We should standardize and  establish best practices based on those already found within missions.
These goals are best served
with a design that enables common user identity models, along with common data, \gls{software}, and infrastructure as services joining systems in a loosely coupled \gls{cyberinfrastructure}. This will drive the the field towards
a more interoperable cross mission \gls{cyberinfrastructure} by design rather than common \gls{API} and
piecemeal translation layers which we currently have.
This will enable developers to reach velocity more rapidly as they move between projects and missions since
they will be  more familiar with the  common development practices and reference architecture.

\subsection{Commodity services and \gls{software} based community architecture} \label{sec:refarc}


The astronomy and astrophysics community community have historically relied on the development and use of bespoke \gls{software} and hardware infrastructure to solve challenges related to the managing and analyzing datasets at a scale that was difficult to find in industry or other scientific domains.
These requirements are no longer unique and we have access to a wealth of open source \gls{software}, commodity hardware, and managed \gls{cloud} services (offered by commercial providers and federally-funded institutions) that are well positioned to meet the needs of astronomers and astrophysicists \cite{2019AAS...23345706M, 2019AAS...23324505B}.
By providing documentation and reference implementations of the “astronomy stack” using these technologies and making it easier for researchers and missions to access \gls{cloud} computing services, we can reduce operations costs, accelerate time to science, and increase the scientific return on Federally-funded research in astronomy and astrophysics.

Such an architecture/system will provide access to new  technologies for improved data \gls{interoperability}. For example to enable a system to recognize transients in multi-observatory data with more than just the photometry.  By housing such data as observing conditions, instrument bias, and even observation proposals within the system, developers can implement common layers at higher levels to provide common access  missions. This can be done without having to specify the complete system for gathering, managing, and formatting the data. Missions can enforce access to either sensitive or proprietary information through role based access control to the data. With a well designed service oriented \Gls{cyberinfrastructure}, cost can be minimized as less
coordination will be needed to implement cross mission services.



\begin{figure}
    \centering
    \includegraphics[width=1.\textwidth]{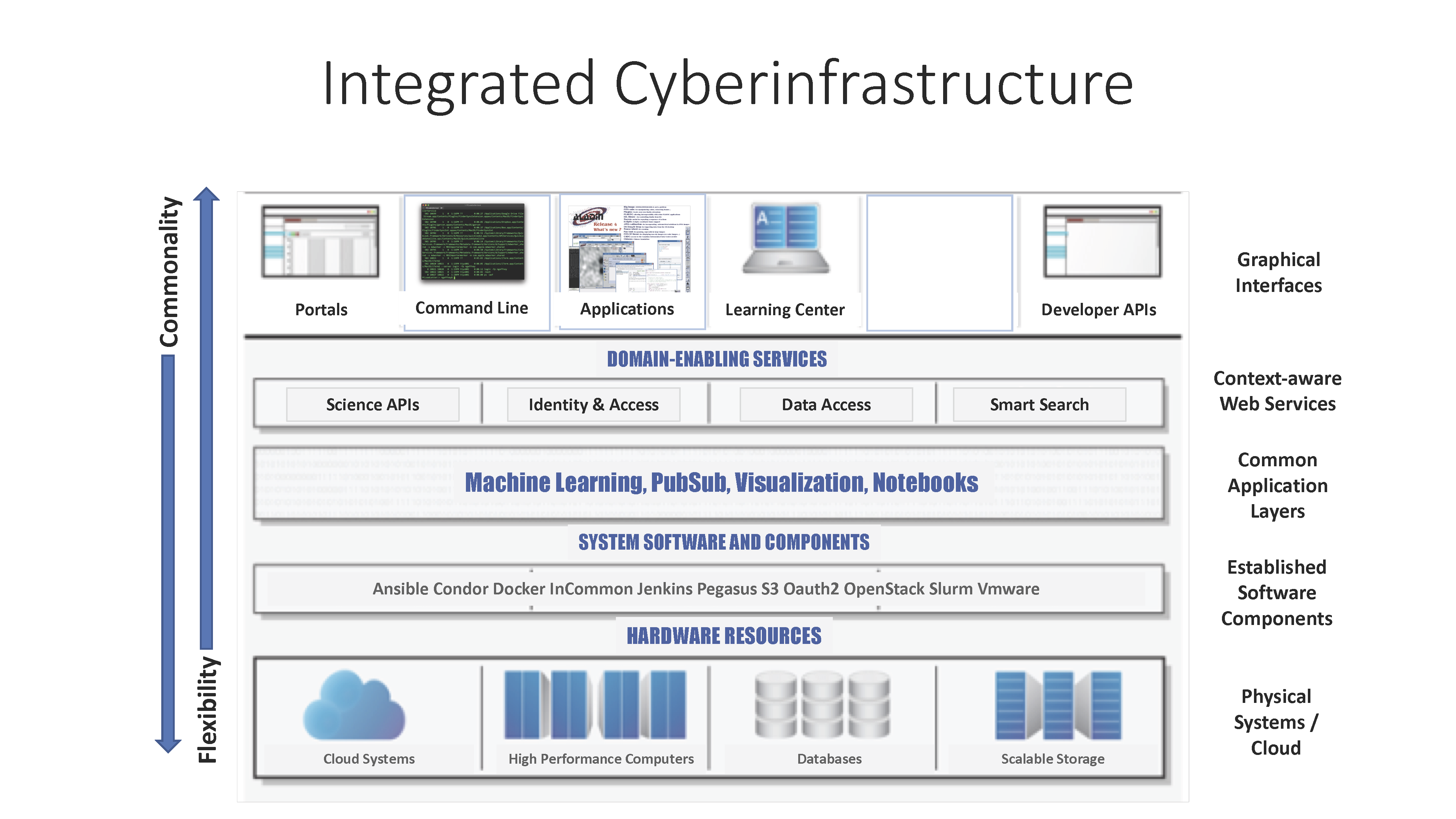}
    \caption{An example a \gls{cyberinfrastructure} built on an Infrastructure as Code design model. Note that while this example does not have astronomy-specific tooling, our recommendations highlight the importance of developing astro-specific layers that are fully accessible to scientists in both  the application  and the graphical interface layers. \figref{fig:CI-LSST} presents an LSST/Astronomy instantiation of this.}
    \label{fig:CI}
\end{figure}\

\nrec{Technology}{Agency, Technologist}{Long}{cbra}{Develop a community wide architecture supporting Science as a Service}{
Agencies should fund the major missions to define and adopt a community wide supported data and compute service architecture with easy to adopt
components leveraging widely adopted infrastructure standards both in the community and in industry.
This "Infrastructure as Code" \citep{morris2016infrastructure} approach lowers the bar to entry
and allows for easier adoption of more standardized services that will enable large-scale
astronomical research in ways that are well demonstrated in plant genomics (CyVerse and Galaxy), natural hazards (Designsafe), and surface water research (Hydroshare).}

Many research communities have accelerated their time to discovery and lowered their cost of integration by adopting a
common community wide architecture that is supported
by multiple data and computational service providers. While
attempts prior to the past decade have been moderately
successful, the current shift in development across
industry to the support of smaller services rather than
monolithic data and compute systems allows for faster
and more cost effective deployment across communities.
By encouraging the definition  and production of  an astronomy focused, community wide reference architecture, perhaps by changing the funding structure, we can being to have a menu of services  easily implementable across service providers.  Design and
support for this infrastructure should be community driven, prioritized and funded, to allow for development of features across missions and science use cases.



Pictured in Figure \ref{fig:CI}  is the structure of a \gls{cyberinfrastructure} (\gls{CI}) that has been
used across multiple fields from plant and animal genomics
(CyVerse) to natural hazards engineering (DesignSafe). This shows the
layers of the \gls{CI} from the interfaces for service access exposed
at multiple levels, the common domain wide enabled services, and a collection of system level components that support the
higher levels of the \gls{CI}.
The lower down the diagram are commodity layers based on well established and supported
components. As one moves up from these layers, more abstraction can be done to
expose these pieces in domain or even question level interfaces. By making these
abstractions, more universal service can be developed that can be applied more globally
across the entirety of the \gls{cyberinfrastructure} as a whole.  An example of this would be
authentication, where each university or agency may provide their own authentication method
but unifying services like CILogin can bring those together to give global spaced
identity for a wide range of users based on disparate authentication systems.
By providing this structure along with a reference architecture of these System Services based on
well supported \gls{software} components, providers are easily able to both deploy and support these common services which enable
cross mission and center \gls{interoperability}. This structure also reflects how this architecture allows for greater reusability as one gets closer to the actual implementation of these
services while supporting greater flexibility and general usability as one works further from the core components.
This service architecture should be based on using standard reusable \gls{software} from many of the established standards developed outside of astronomy (e.g. common authentication mechanisms such as CILogin, standard data and metadata management systems).  Standard \gls{API} interfaces should also be used to expose these components to higher level \gls{API}s. Data
formatting and metadata structure can be exposed at the service level, allowing for
more data and metadata reuse.

\begin{figure}
    \centering
    \includegraphics[width=1.\textwidth]{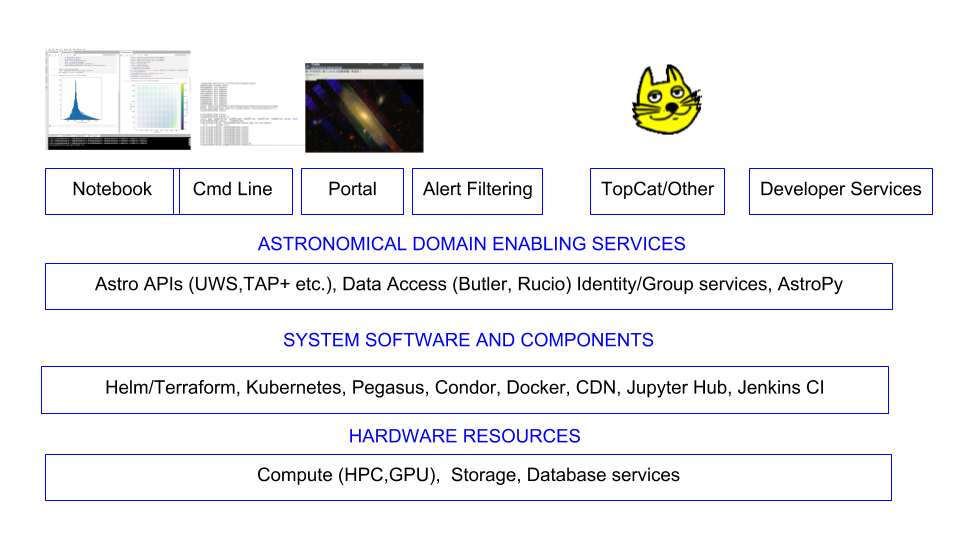}
    \caption{An example \gls{LSST}  \gls{cyberinfrastructure} built analogous to the \gls{CI} model shown in \figref{fig:CI}.}
    \label{fig:CI-LSST}
\end{figure}\

Such an architecture should be developed in a \gls{cloud} and
vendor agnostic manner.  When needed, vendor specific \gls{software}
or \gls{cloud} service can be integrated by a mission, but by isolating them in the \gls{cyberinfrastructure}  at the lowest level, their potential impact on the overall system is minimized. When possible, standard interfaces should be used to abstract out
these differences (e.g. standard object store access like Amazon \gls{S3}, Google Cloud Storage, Microsoft Azure Cloud Storage, standard database interfaces like \gls{ODBC}) and should be reflected in the reference architecture documentation. Where practical, computational environment abstraction layers such as
container technologies (e.g. \gls{Docker}) should be used to associate each
applications
computational requirements to the application rather than having to
enforce upgrades and updates across the complete infrastructure. Where
specific hardware environments are required (e.g. Google's Tensor Processors or NVIDIA GPUs),
it must be
well-understood programming frameworks for access (e.g. Tensorflow or CUDA)
to allow for simpler migration to future or
separate vendor's systems.


Many of the advances in the current data revolution have come about from the
broad adoption of commodity hardware and \gls{software} services being applied in
domain agnostic ways. While over a decade ago, astronomy was one of the first domains to
explore these technologies, the current momentum in the area is driven outside
of the astronomical field. Current technologies like Spark, Cassandra, Kubernetes, and
MongoDB were all created to support large data problems outside of astronomy but are
finding support in the astronomical community in a piecemeal manner. By shifting
the focus from local to a more distributed \gls{cyberinfrastructure}, such new technologies could be
implemented and leveraged much quicker than if each center had to support and
integrate their own solution, Further, by embracing and adopting a more commodity base infrastructure will allow current and future projects to choose and experiment with hardware alternatives such as \Gls{TPU}s, \Gls{FPGA}s, or Quantum computing as they become more commonplace or to integrate newer architectures that best suited to the problem.

There is a current tension or incompatibility with the direction of High Performance Computing \gls{HPC} to many more cores with not much more memory and large image processing which requires more throughput. Historically, systems designed for \gls{HPC} were
more suited to simulation as opposed to the embarrassingly parallel yet often memory intensive High Throughput Computing \gls{HTC}.
While \gls{HTC} does not get as much attention, it has been key to missions such as the Human Genome project,
\Gls{LIGO}, and \Gls{SDSS}.
While some are moving to bridge this gap (some in support of missions such as \gls{LIGO} and others because simulated data analysis is becoming
as complex as observational data analysis), agencies should continue encourage
and fund national computing centers to address the needs of both communities. Further, they should
enable simulated datasets to coexist within the infrastructure with observational datasets to support the full cycle of astronomical discovery. Finally, they should
formally adopt support for mission long computational support at these facilities
for all major missions.

We note the general idea here especially on the importance of \gls{cloud}  are compatible with  Rob Pike's thoughts\footnote{\url{https://drive.google.com/open?id=1kYsavh900I2o6z1lPfjFFXamyBkoEdCW}}.

Where ever possible, review of \gls{cyberinfrastructure} driven proposals should be ranked on both their immediate
impact to the field as well as their ability to sustain such impact through several technology and vendor cycles. Proposals should also be ranked based on their reuse and/or integration into the overall \gls{cyberinfrastructure} developed across the astronomical community.


\subsubsection{Identity and Access Management (\gls{IAM})}\label{sec:iam}
User identity is key to all data systems to date. In the
past, each mission has had its own \gls{IAM} system to provide data access
for embargoed or otherwise restricted data access and for accessing
services. With a more global \gls{CI}, this problem must be further abstracted to allow for the notion of identity of a user (aka authentication) and
permissions (aka authorization) for services and data. By separating
these pieces, identity can be brought from multiple sources (CI-Login,
OAuth, ORCID) while the permissions can be enforced by each provider.
All aspects of this \gls{CI} must embrace role based authorization for both data and services in a federated manor so that data and services
can be effectively orchestrated while not impacting site or mission
specific access restrictions. While the authentication may be global, each provider will enforce their own access roles for all users. As data analysis
moves into the petabyte and even exabyte scale across multiple missions
and multiple computational environments, it will be paramount to
for all members of the \gls{cyberinfrastructure} to share a common \gls{IAM}
infrastructure to allow for federated access controls for both
data and computational services.

Proposals and missions should be ranked on their adoption of
such an \gls{IAM} for data and for \gls{software} services. Further, by
creating a common \gls{CI} layer for identity, missions and smaller
services will be able to adopt and adapt the common system to
their needs, thus saving development costs for implementing and
finally supporting their own system.


\subsection{Data co-location and creating data lakes}\label{sec:coloc}

 \nrec{Technology}{Technologist}{Medium}{coloc}{Enable new scales of research through data co-location} {We must enable the co-location of data thus enabling large scale cross mission research. While some
 research are well supported using specific services, ones which require comingling data to produce new data products or results should be able to schedule access to co-located data in ways similar to acquiring new data.}

 Astronomy archives, especially in the USA, are somewhat fragmented. Though \gls{TB} scale permanently co-locating all data n one or more centers is technically possible it is probably not desirable nor socially possible.
In the \gls{PB} data era permanent co-location becomes less feasible, yet  researchers will require data lakes or reservoirs to house massive datasets and allow computations to be done across multiple missions and epochs
of data at scale. The data lake concept, where data from multiple missions and schemes
are temporarily co-located so as to allow codes to more tightly integrate with multiple data sources,
is the most attractive for researchers (who have become accustom to immediate data
access), the idea of a reservoir that can be filled and drained of data based on
the demand of users is one that will need to be explored due to the economical
viability of any one institution housing all the data.

Collocation is more than just \gls{interoperability} but  will also mean  generating
new formats  e.g. \gls{parquet} files (e.g.) to make dynamic
data frames and data services as demands change.  It will also mean generating
new data products across missions which are equally valuable for both
preservation and publication. Thus the lake is more than a simple pool of
storage, but should be operated similar to other key infrastructures in
the observational astronomical infrastructure.

Software and technologies change on timescales faster than a
decade (see also \secref{sec:swalive}) and  data centers need  to be agile enough to keep up.  One approach is to create
interoperability interfaces to allow data to be pulled
or pushed from managed repositories to dynamic data
lake environments where users can produce their own custom subsets mixing the available
datasets.
Data movement will not be simple (see also \secref{sec:net}) and to not be prohibitive
specific infrastructure would need to be supported and evolved.

While the immediately obvious argument for co-locating data is the potential for scientifically rich co-processing of heterogeneous data holdings (e.g. Euclid, \gls{WFIRST} and \gls{LSST}), the advantages do not end there. Co-locating large data holdings on a common commodity computing platform enables bring-your-code-to-the-data capabilities (sometimes referred to as Science Platforms or server-side analytics) to enable co-analysis of data from a single service without necessitating data transfer. For example, a single Jupyter notebook can present an runnable analysis drawing on separate datasets (e.g. a multi-wavelength analysis of a class of objects). Furthermore, co-locating data holdings allows the co-location \emph{and sharing} of services accessing those data holdings. That not only includes the possibility of collaboration in sharing a single \gls{API} service between data publishers, but also reducing the development burden of infrastructural services (for example, documentation infrastructure could be shared with multiple missions, migration paths to new operating systems are easier, etc). In an era where Infrastructure as Code engineering paradigms represent emerging best practice,
re-using the code that underwrites common astronomical services because they are developed within a common underlying infrastructure (such as \gls{Docker} containers orchestrated by Kubernetes) provides an avenue for fruitful ongoing collaboration between data publishers and better value for money for science infrastructure dollars.

\subsubsection{Network layer}\label{sec:net}
One area where Infrastructure as Code does not work is networking.  While data transfers
can be better optimized depending on the nature of the data being transferred, network
as a service will require optimizations often at the point to point level. Tools like
PerfSonar and others from Internet2 can help optimize connections. But these
have been used to optimize research done at the terabyte scale today. Where as in the
\gls{TB} scale it was possible sometimes to move the compute to the data (e.g. \gls{SDSS}), discovery often comes from the \gls{TB} scale data lakes. To meet the
demand for petabyte scale data motion as a service in a terabit network age, systems to support Just In Time data
delivery are needed for any large scale data collocation  and must be a part
of the support for the overall infrastructure of repository/cloud/data center
collaborations.  Computation on large datasets may need to be scheduled no
different than any other instrument used in observational astronomy.

We recommend funding projects in astronomy (and in other research domains) to
create the data management and migration layers and best practices that will
enable these new forms of observation. We also recommend that proposals show how they
will develop and sustain such services over the course of the mission and beyond.

\subsubsection{Vendor freedom}\label{sec:lockin}
When co-location of data holding and services on a \gls{cloud} platform is discussed, inevitable there are concerns raised about "vendor lock-in".
These objections are often rooted in a misunderstanding of the nature of these services. These services often themselves share identical or nearly identical programming interfaces for application developers. For example, all major Cloud Storage Services such as Amazon S3, Google Cloud Storage, and Microsoft Azure Cloud Storage share very similar interfaces. Furthermore, the commercial landscape is designed around low barriers to change: Google Cloud Platform, Amazon Web Services, and Microsoft Azure make it possible for customers to switch their services from one Cloud provider to another.
Moreover all these platforms drive a service-oriented architecture that inherently results in more portable systems. In any case if one is concerned about vendor lock-in, in-house data center infrastructures are the worst possible choice: the inevitably lead to infrastructure-specific choices that are poor candidates for evolution, and they typically lack the effort to support ongoing refreshing of the technical stack thus creating on-going support burdens and a change-averse culture.
For example, \gls{LSST} is a project that will go into operations in 2022, and yet \gls{LSST} \gls{DM}  are frequently called on to support CentOS 6, an operating system released in 2011, because that is the only \gls{OS} some university in-house clusters support for researchers.

\subsection{Operations}\label{sec:ops}
\nrec{Technology}{Agency}{Short}{fundops}{ Enable support for full mission life cycle including long-term data products } {Agencies should revisit the model separating funding and
requirements for development and operations of large scale missions for which the data and services are key deliverables as well as steel and concrete.  Such services are under
continual development and integration and, in the current environment, can not simply be
maintained in the same way physical facilities are.}

A more operations oriented view of construction by funding organizations would lead to facilities which are more cost effective to run.
Recognizing that \gls{software}  and \gls{cyberinfrastructure} development work is distinct  from concrete and physical facilities  would also help to make more maintainable and agile \gls{cyberinfrastructure}.
Current MREFCs funding splits construction from operations  thus limiting support on ongoing work for  \gls{cyberinfrastructure},  there is little incentive in construction to build easy to operate systems.\footnote{This is also true for physical facilities e.g.  autonomous operation is usually not a requirement.}
If we blur the line between construction and operations for \gls{cyberinfrastructure} the issue becomes more one of long term support.

\nrec{Technology}{Technologist, Agency, Astronomer}{Short}{opspractices}{Improve long-term \gls{software} and service support}{Funding should  support repositories not just for code and data, but for computational environments. Use of proven standards in the wider research community for sharing and discovering runtime-ready \gls{software} using \gls{software} \gls{container} environments like \gls{Docker} and \gls{Docker}Hub but with domain specific curation
(e.g. BioContainers) is crutical\footnote{Critical and crucial \url{https://www.urbandictionary.com/define.php?term=crutical}} for both broader impacts of \gls{software} products and result
reproducibility as  compute environments continue to rapidly evolve with low emphasis given to
backward compatibility. }

Long term support could be improved by requiring proposals to state how code reuse and common practices will be used. Looking for  proposals which aim to push to the Code and
    \gls{API} communities (AstroGit? AstroContainers? AstroHub) and which aim to build on common
    \gls{software} development practices. Of course we should also foster development of those best practices
 based on best practices outside astronomy.
Concretely proposals could have  a line item for making \gls{software} reusable in funding budgets and funding agencies should see that as a good thing and try to develop metrics for success in the area.

We must also consider how to move to a more self service architecture in astronomy  such as GitOps \citep{Limoncelli:2018:GPM:3236386.3237207} - that requires some rigor but establishment of adhered to best practices would be a start.

Needless to say all of the code should be available under open licensing such as Apache (\gls{APL}) or Gnu (\gls{GPL}) public license.

\nrec{Technology}{Agency}{Medium}{cmdeploy}{Fund cross-mission deployment} {
    Missions that develop \gls{software} that can and should be adaptable to other common goals in other missions should be funded to develop and support cross-mission resources as part of this and other cyberinfrastuctures.
    }
Past examples
    of this success can be found as far back as the \gls{IRAF} environment and include now significantly
    more broadly adopted numpy, the cost of development and support have been paid back multifold
    across many funded projects and missions which did not have to develop their own versions
    of \gls{software}.

Deployability  is part of the problem but service oriented architectures are and will remain at the forefront for at least the next decade.
So we should now be thinking more of \gls{software} as a service and defining infrastructure as a service.
This would all funding agencies to push us more toward commodity compute and infrastructure services thus concentrating efforts on the astronomy problems at hand rather than the computer science problems.


Funding agencies could also  favor proposals that use/leverage existing \gls{software} solution, it may take time but this would be a positive fundamental change in astronomy \gls{software}.


\subsection{Sustainability and effectiveness}
\label{sec:tech_sust}

An architecture as laid out in \secref{sec:refarc} gives us a framework to start a sustainable \gls{software} development in astronomy.
Software sustainability is a complex issue, with mixed results in astronomy --- see \secref{ssec:soft_sustain} for more on this complexity.
The approach laid out here, however, presents greater challenges than ever before in sustainability, as larger datasets and more complex infrastructure means that reinventing the wheel will become ever more costly.

By providing a reference architecture we can allow for more openness, collaboration, and when necessary, competition in each component.
While competition and alternatives can be useful in a controlled manner, i.e. funding two approaches for a specific component, by providing a reference architecture this can be adopted only when needed without interfering with other layers. We should also consider that this architecture may be good for a decade at which point it also should be revisited - such a refresh should be built in to our thinking from the start. This is critical for sustainability because sustainability is much more challenging if the architecture does not keep up with the technology, as more and more ``hacks'' become necessary until the house of cards comes toppling down.

The Astropy project has been successfully fostering a community approach to the user-facing end of Python astronomy. They do face challenges for funding and are beginning to tackle some management issue for an organically grown organisation. This has been successfully because they have worked hard to join the zeitgeist of open \gls{software} and have dedicated individuals who believe this is a useful project - and many users who agree. The project also produces useful tools.
The role of the Astropy project in the ecosystem can be misunderstood i.e. it is not meant to be a data management system or a replacement for a mission-specific toolchain, rather it is a toolbox that is accessible to astronomers for doing many of the tasks they want to use or understand. While the re-use and contribution to these tools by missions or observatories is desirable (see \secref{sec:software}), without a clear understanding of where the components lie in a larger architecture, it is almost impossible for astronomers and projects to understand how they can fit such components into their system. A reference architecture can thus help define where projects like this belong and how they fit with other efforts. For example, \gls{LSST} will not process data using \texttt{astropy} but will make their data products accessible and usable with particular Astropy project interfaces and tools. It has taken both projects a while to understand this delineation because  neither \gls{LSST} nor \texttt{AstroPy}  had a clear reference model to work with.

\section{Software \Contact{Arfon Smith, Erik Tollerud}}\label{sec:software}

\Contributors{Arfon Smith <\mail{arfon@stsci.edu}>, Erik Tollerud <\mail{etollerud@stsci.edu}>, Kelle Cruz, Stuart Mumford, Nirav Merchant, Gautham Narayan, Alex Drlica-Wagner}
\bigskip

\noindent\textbf{Note:}
If you have come directly to this chapter we suggest you please read at least the Introduction in \secref{sec:intro} before delving further.
\newline

In the Petabyte era, all projects are software projects --- that is, sophisticated software is necessary throughout the system for essentially any scientific output at this scale. That said, the term \gls{software} can mean many things to many people. Other Sections (e.g. \secref{sec:science}, \secref{sec:data}, \secref{sec:algorithms}) discuss the \emph{content} of this software, and the software ``infrastructure'' components and how they fit together are discussed in more detail in \secref{sec:tech}. Here, by contrast, our focus is on the process by which software distributed to and used by the astronomy community is built and funded. We particularly focus on \gls{community software} as it is the most relevant for the community to be involved in and for funding agencies to support. Note that that while clearly critical to the process of software development, relevant career and workforce issues are discussed separately in  \secref{sec:iwft:software}.


\subsection{Recommendations}

\nrec{Software}{Agency}{Long}{swmoney1}{Funding for \gls{software} development in existing grant programs}{Software that enables science should be allowable as a sole deliverable for all existing funding programs (e.g. \gls{NSF} \gls{AAG}, \gls{NASA ROSES}, postdoctoral prize fellowships). It should not be necessarily coupled to a specific science effort, as long as the \gls{software} is of demonstrable use to the scientific community.}

\nrec{Software}{Agency, Astronomer, Technologist}{Medium}{swmoney2}{Funding for sustaining core astronomical ``community infrastructure'' projects}{Funding agencies and the community as a whole should support funding of domain-specific community-developed \gls{software} projects e.g. Astropy project, SunPy.  Such projects should be recognized as vital infrastructure and placed on an equal footing to physical facilities such as national observatories.  This support should be available for domain-specific \gls{software}, rather than funding being primarily tied to interdisciplinary applicability. It should also be allowed to fund community-development efforts in addition to actual code development.}

\nrec{Software}{Agency, Manager, Astronomer}{Short}{swecoplan}{Cultivating a sustainable research \gls{software} ecosystem}{Funding agencies should include as part of their review criteria for all astronomy grant programs: 1) A plan for how \gls{software} (not just data) will be managed to support the science of the grant, 2) How proposed \gls{software} development fits into and supports the wider ecosystem of available tools, and 3) Favor programs that propose developing \gls{community software} as part of their funded activities. These same goals and considerations should also be considered and acted on by the broader astronomy science community when e.g. working as grant review panelists.}

Note that cultivating a research \gls{software} \emph{workforce} is critical to all of the above. Hence, while not detailed in this Section, the recommendations of \secref{sec:iwft} are also as critical to the discussion in this section as the above.




\subsection{Why shared \gls{software} matters, today and in the next decade.}

In this chapter, we argue that the Petabyte era of discovery in astronomy means that the role of \gls{software} is increasingly important and that well-organized, well-maintained \gls{software} serves to shallow the learning curve, enable scientific investigation, and lends confidence to scientific results.
To set scope, though, we emphasize this mainly applies in the context of shared \gls{software}.
That is, the ``throw away'' analysis script written, say, by a graduate student when writing their thesis and never shared with anyone else does not count for this discussion.  However, that changes when the same student shares that script with their collaborators, makes it available online, or contributes it to The Astropy Project, rOpenSci, or another open \gls{community software} resource. Such an act makes the software part of a community process, and the astronomy community is the target of this discussion.  Hence, in this chapter we are focused on \emph{shared} software.


\subsubsection{Software is everywhere}

\begin{quote}
    ``\textbf{Software is a central part of modern scientific discovery}. Software turns a theoretical model into quantitative predictions; \gls{software} controls an experiment; and \gls{software} extracts from raw data evidence supporting or rejecting a theory'' - Gaël Varoquaux, scikit-learn\footnote{\url{https://scikit-learn.org}} creator, \citep{varoquaux, scikit-learn}.
\end{quote}

Software is an integral, and growing part of the scientific endeavor: It is responsible for driving the control systems of instruments, the operation of surveys, the processing of raw data products, the extraction of physical parameters, and the theoretical modeling of physical systems, \gls{software} is critical to all parts of modern computational research. Indeed, `Software is eating the world' \citep{andreesen}. This reality is well-recognized by the scientific community: In a survey carried out in 2009, more than 2000 scientists reported that \gls{software} was either important or very important to their research, and that it would be impractical for them to carry out their research without it \citep{Hannay2009}.

The rapid increase in the size and complexity of astronomical experiments and the data they produce has led to an increasing demand for astronomical software. An illustration of this point is the \gls{LSST} project and their allocation of 25\% of the construction budget (\$187M) for data management \gls{software}, infrastructure, and services\footnote{\url{https://www.nsf.gov/about/budget/fy2018/pdf/30b_fy2018.pdf}}.

Over the last decade, in a large part driven by broader changes in the cultural `norms' of modern \gls{software} development and a shift towards \gls{open source software} being `the new normal' \citep{opensourcedotcom, forbes}, the astronomical \gls{software} environment has changed rapidly: Large experimental projects (such as \gls{LSST}, \gls{JWST}, \gls{DESI}, \gls{DKIST}) are writing extensive code bases and releasing these tools as open source software\footnote{\url{http://github.com/lsst}}\footnote{\url{http://github.com/spacetelescope}}\footnote{\url{http://github.com/desihub}}\footnote{\url{https://github.com/DKISTDC}}. At the same time, individuals are becoming increasingly likely to distribute and share their code broadly with the astronomical community and mechanisms for publishing these \gls{software} products have expanded as a result \citep{aassoftware, githubzenodo, Smith2018, astroandcomp}. In this data intensive future, where \gls{software} permeates scientific investigation, it is critical that the contributions of \gls{software} developers are recognized and that individuals are provided with the necessary resources to succeed.

\subsubsection{Software encodes knowledge}

As datasets become larger and our analysis methods more sophisticated, an increasing fraction of the scholarly method is expressed in software. This presents opportunities and challenges. One potential opportunity is that the `centralization' of astronomy (i.e. the trend towards smaller numbers of large facilities, often with open datasets) means that any \gls{software} built (and shared) leveraging these facilities has a higher reuse potential. A major potential risk, identified by others \citep{Donoho2009, stodden2010}, is that as the fraction of our research method is captured in \gls{software}, if this \gls{software} isn't shared (e.g. as open source), reviewed, or tested, the \gls{reproducibility} of our science is increasingly at risk.

\subsubsection{Software for reproducibility}
As projects become increasingly complex, ever more discrete \gls{software} components are combined to produce analyses and data products. However, despite this complexity, much of the code being used is not documented, let alone complete with unit tests that can validate performance. These shortcomings can have real-world consequences, as illustrated by the failed Mars Climate Orbiter mission, where \gls{software} calculations were carried out assuming metric units, but navigation \gls{software} was programmed assuming imperial units, leading to a premature and fiery end to the mission in the Martian atmosphere. While this is an extreme case, it is an illustrative bounding case for more subtle problems in analyses that lead to biases which are not detected.  Such problems are surprisingly common even in the computational science literature \citep{Collberg16}, much less more applied fields like astronomy.  While progress has been made in developing technologies to improve this (e.g. easy and widely available \gls{software} repositories like GitHub, containerization technologies like \gls{Docker}, etc), many of these technologies are still aimed at early-adopters in science rather than the mainstream.

\subsection{Progress in the last the decade}

Many of the issues highlighted in this chapter are not new. In particular, we highlight a white paper from the Astro2010 decadal review with a similar scope: \citet{weiner09}. That paper discussed areas of concern and specific recommendations, some of which have improved materially, while others have seen little progress. We discuss the recommendations of that paper here to provide a historical context and guidance for the future.

\begin{enumerate}
    \item \citet{weiner09}: \emph{``create a open central repository location at which authors can release \gls{software} and documentation''}.  Enormous progress in this area has been achieved in the last decade. \gls{open source software} repositories, chief among them GitHub\footnote{\url{https://github.com}}, have become a defacto standard for storing \gls{software} in astronomy. The wider adoption of Python has improved the  packaging and release process due to the Python Package Index\footnote{\url{https://pypi.python.org}} and the ecosystem of easy-to-host documentation tools that support Python, like Sphinx\footnote{\url{http://www.sphinx-doc.org}} and ReadTheDocs\footnote{\url{https://readthedocs.org/}}. While these are not a perfect solution for some languages and science domains, the presence of a much larger and better-funded user base (open source industry \gls{software}) has made them stable enough to be adopted for astronomy's use and can likely continue to do so for the foreseeable future.
    \item \citet{weiner09}: \emph{``Software release should be an integral and funded part of astronomical projects''}. Progress in this area has been mixed. While large efforts for this decade like \gls{LSST}, \gls{JWST}, \gls{DESI} or \gls{DKIST} have large first-class \gls{software} components, many smaller projects or individual grant-level efforts continue to treat maintainable or reproducible \gls{software} as an afterthought to be dealt with in whatever time is left over by graduate students or postdocs rather than a necessary part of the scientific endeavour. While funding agencies like the \gls{NSF}, \gls{DOE} and \gls{NASA} have required data management plans, there has been less progress on establishing firm requirements or expectations of sustainable \gls{software} (although a recent NASA-driven consideration of these issues is available in \citealt{nasaosp}).
     \item \citet{weiner09}: \emph{``Software release should become an integral part of the publication process.'' and ``The barriers to publication of methods and descriptive papers should be lower.''}. Considerable progress has been made in this area.  The \gls{AAS} journals now allow software-only publications on equal footing with more traditional science publications \citep{aassoftware}, and other major astronomy journals like A\&A and \gls{PASP} do as well.  New approaches to publication like the Journal of open source Software\footnote{\url{http://joss.theoj.org}} \citep{Smith2018} or the Astrophysics Source Code Library\footnote{\url{http://ascl.net}} are now providing alternate ways to publish \gls{software} that are indexed in \gls{ADS}.  Software archives like Zenodo\footnote{\url{https://zenodo.org}} now connect with GitHub to make publication of \gls{software} via \gls{DOI} almost frictionless \citep{githubzenodo}.  While there are still challenges in identifying how \gls{software} citation should work in these areas, tangible progress and recommendation is being made \citep{Smith2016}.  The ``cultural'' elements of ensuring these publications are viewed with the same level of value as other publications may also be improving, although concrete data in this area is lacking. While somewhat less progress has been made in ensuring open \gls{software} is a truly integral part of publication, the same resources noted above have made it much easier to preserve \gls{software} long-term.  More challenging is preserving the \emph{environment} \gls{software} has been run in.  While technologies like \gls{Docker} or virtualization provide a possible path, they have not been adopted widely across the community thus far, and represent a possible major area of development for the 2020s.
    \item \citet{weiner09}: {\emph{``Astronomical programming, statistics and data analysis should be an integral part of the curriculum''} and \emph{``encourage interdisciplinary cooperation''}}.  While some progress has been made in this area, there are many challenges remaining.  We defer further discussion of this to Sections \secref{sec:iwft}, \secref{sec:epo}, and \secref{sec:algorithms}.

    \item \citet{weiner09}: \emph{``more opportunities to fund grass-roots \gls{software} projects of use to the wider community''}. While such projects have grown remarkably in the last decade (see \secref{ssec:opendev}, major challenges still remain in \emph{funding} such projects in a sustainable manner, and these form the core of some of our recommendations.
    \item \citet{weiner09}: \emph{``institutional support for science programs that attract and support talented scientists who generate \gls{software} for public release.''}. Some of the elements of this recommendation have grown with the advent of ``Big Data'' and ``Data Science'' academic positions in astronomy.  There has also been a growing recognition of the importance of research-oriented \gls{software} positions, particularly in Europe \citep[e.g.][]{rse}.  However, there are very few viable pathways for researchers who develop \gls{software} of broad use as part of their research program if it is not considered a ``hot'' field.  Because, as this book demonstrates, there are likely to be \emph{more} areas where deep \gls{software} expertise is critical to science in the coming decade, the need for the field to nurture such career paths will only become more acute.  Hence this is also a key element of our recommendations.
\end{enumerate}

There is one final distinction to be highlighted relative to the last decade: it is clear that \gls{software} has become more mission-critical than in the past. As the other chapters of this book highlight, in the coming decade(s) large-scale science will require larger and more complex software.  These generic concerns about \gls{software} development are therefore multiplied across the deeper layers of \gls{software}, making all the issues more broadly applicable. The urgency in addressing these issues will only grow in the coming decade.

\subsection{Community \gls{software} as a force multiplier}
\label{ssec:forcem}
Collaboratively-developed \gls{community software} has an increasing large impact throughout the astronomy community.  For example, the whole scientific \gls{software} ecosystem in Python \citep[the most popular language in astronomy][]{momcheva15} is built on community-developed \gls{software} like NumPy \citep{numpy}, SciPy \citep{scipy}, Matplotlib \citep{Hunter:2007}, or other parts of the so-called ``NumFOCUS Stack''. More domain-specific projects such as Astropy project\citep{astropy1, astropy2} and SunPy \citep{sunpy} capture the expertise of a broad range of astronomers, and have a wealth of features that cannot be reproduced by solitary researchers.  While the mere existence of such \gls{software} open to all to use are immediately apparent, there are several ancillary benefits to such \gls{community software} efforts:
\begin{itemize}
    \item The more the community participates, the more the project will reflect their specific needs and applications, even if it is built on a more general framework.
    \item The code is typically inspected by more people, and many eyes make all bugs shallow \citep[i.e. code problems and their solutions will be quickly found][]{raymond01}.
    \item There is usually more documentation available because of the free energy to specialize on such tools, and a larger base to help support new users.
    \item It is easier to train scientists to help produce professional-quality \gls{software} if they are supported by a core of professional engineers. Community projects provide a larger-scale social understanding of how that interaction can happen.
    \item These projects speed up the cycle of science by providing useful implementations for common tasks, freeing up researchers to work on their specific science.
    \item When built as part of an underlying broader ecosystem, \gls{community software} often gains the direct benefit of contributions ``upstream'' e.g. improvements in core math libraries made by computer scientists can flow down to astronomy without any direct effort in astronomy.
\end{itemize}

Together, these factors mean that the impact of code developed by a community is multiplied by further contributions from other sources to the same ecosystem.

We note that the community developed \gls{software} need not strictly be open source, though the majority of these projects are. The benefits of community development extend to both open and closed source projects, the primary difference being that the potential size of an open project is by definition larger than a closed one, and most of the above scale with community size.

\subsubsection{Open development/Open collaboration}
\label{ssec:opendev}

While a substantial fraction of software in Astronomy is now \gls{open source software}, and has been for decades, a major development in recent years has been the growth of \gls{open development}. This form of collaboration software development accepts and in many cases depends wholly on contributions from the wider community to the software project. Development of the code and discussion around that code is conducted in the open using industry-standard platforms like GitHub or GitLab, and in most cases policy discussions and decisions also occur in the open, or example on a public internet mailing list.  The chief examples of projects like this in astronomy are The Astropy project and SunPy.

This kind of development model is not limited to astronomy projects, there are many examples of large scale \gls{software} projects which are entirely developed in the open, the largest example of which is the Linux kernel. Developing \gls{software} in this way introduces technical and sociological challenges, which have been met by \gls{DVCS} tools such as \gls{git}, online collaboration tools such as GitHub that enable workflows which scale to many hundreds or thousands of contributors, and the hard work of organizers and code reviewers to set up and maintain a positive culture that enables contributions to continue.

These kind of open collaborations enable many different stakeholders (both astronomer-users and dedicated developers) to collaborate on a \gls{software} project, often from a diverse set of perspectives. While this is possible with non-open developed \gls{community software}, it is often much harder because it requires an added layer of communication between ``users'' and ``developers'', while in \gls{open development} these are the same community. This makes the \gls{software} more valuable to both the contributors and the community more than the sum of the individual contributions, as it reflects the needs of the many rather than the one. It also means more work can be done with less funding, because the efforts of individual contributors are pooled into a ``neutral'' space that can arbitrate via the community process. Moreover, the open nature of the collaboration means that stakeholders have the ability to drive the direction and priorities of the project simply by contributing to it. Because many of these stakeholders are the users themselves, it also can serve to optimize the applicability-to-effort ratio.

\subsection{Community \gls{software} problems and solutions}

With the above in mind, there is incongruity between the increasing importance of \gls{community software}, and the funding available for such projects. In particular, the future of many widely used projects that are effective force-multipliers, including \texttt{astropy}  and services such as \url{astrometry.net}, are uncertain. These major community projects are generally unfunded despite the vital role they play for astrophysics as a whole. While many feature ``in-kind'' contributions from user missions (as discussed above), such support depends on the vagaries of mission priorities rather than the needs of the community itself (as discussed below).

Hence, the benefits outlined above cannot be realized if such efforts are not supported by funding agencies, large missions, and indeed the astronomical community as a whole. Currently incentives are not in place to encourage community efforts: indeed in some cases such \gls{software} development is either not allowed by a grant program, or tacked on as an afterthought. (``Oh, we'll probably have my grad student build that reduction pipeline on the way to their thesis.'') Where \gls{software} grant programs do exist, they often focus on building specific applications into interdisciplinary tools (e.g. \gls{NSF} \gls{CSSI} and \gls{DIBBs}), rather than applying general \gls{software} to specific domains. They also as a rule do not emphasize \emph{community-building} elements like contribution policy documents, documentation of user workflows, or community coordination. Hence, while specific recommendations of what platforms for development are useful are not likely to be relevant in 10 years (and indeed are often counter-productive - see \secref{sec:swalive}), our recommendations focus on incentives for pro-social behavior by missions and individuals. This will be critical to keeping up with the ever more software-rich Petabyte era, and this is precisely what the recommendations of this chapter aim to do.

\subsection{Software is alive}\label{sec:swalive} 

\label{ssec:sw_alive}

\begin{quote}
    ``This open source stuff is free. But it's free like a puppy. It takes years of care and feeding.'' - Scott Hanselman on the death of nDoc \citep{ndoc}
\end{quote}

The grant-funding model for academia fosters a picture of all work as limited to a fixed time horizon, shared astronomical \gls{software} often lives as long as it is useful.  This can be far longer than any individual researcher or developer, and as a result the \gls{software} takes on a life of its own.  Like any living thing, however, this \gls{software} will not survive without proper care and feeding, and without evolving to adapt to continually changing environment.

\subsubsection{The \gls{software} stack is always changing. We need to be adaptable.} \label{sec:stack}

Sustainability is a necessary but not sufficient condition for \gls{software} to survive. Even with maintenance, the entire \gls{software} ecosystem is constantly evolving. A clear example is Python replacing \gls{IDL} as the most popular programming language within astronomy \citep{momcheva15}, despite many of the elements of the \gls{IDL} Astronomy Library being maintained. Similarly, many of the features of \gls{IRAF} are now being provided by widely used community projects such as Astropy project, despite the long history of \gls{IRAF}. New \gls{software} like this generally evolves because they can tackle problems that were not addressed previously, either by making the coding easier or taking advantage of other developments in the wider technical world (discussed more above). For example, resasons for the change from \gls{IDL} and \gls{IRAF} to Python are the lack of license fees, the extensive open source ecosystem of libraries for scientific computing, and the easier learning curve of the latter (due to more broad usage).

However,  the disruption caused by the evolving \gls{software}  ecosystem can be disruptive because it comes at the cost of requiring significant retraining and refactoring. In this way, the need to be adaptable to changing developments in \gls{software} can appear to be in tension with the need for well-validated \gls{software} for research. There is indeed always a cost-benefit analysis for changing technologies that most include this concern as much as the benefits that may result.  But consideration must be made that this disruption can be ameliorated by continuing education programs for researchers at all levels.  Examples include \gls{AAS} workshops, introducing astronomers to the up and coming \gls{software} projects and to highlight long-term trends, such as which projects are growing in support vs which are now largely unmaintained.  There are further more focused recommendations in \secref{sec:iwft} for keeping the community on top of such changes. Hence, the disruption caused by the continuous evolution of the \gls{software} stack should not be feared, but rather welcome for its potential to improve our own research.

\subsubsection{Software needs to be sustainable}

\label{ssec:soft_sustain}

Any \gls{software} that is meant to be used more than once requires maintenance. Data sets change (or grow to Petabyte scale), bugs are discovered, computer architectures change, and users change their understanding of the intent of the software. This leads to the concept of \gls{software} sustainability: practices both within the \gls{software} itself and of those who develop it that make it practical to maintain the \gls{software} for an arbitrarily long time.  For astronomy \gls{software} to be sustainable~\citep{wssspe51, wilson14}, it should:

\begin{enumerate}
    \item Be both  testable and tested (i.e. it is correct and that correctness can be checked by anyone).
    \item Be readable and useable by multiple people (i.e.  it can  evolve to fulfill its intent over time as development and scientific conditions change).
    \item Have a viable pathway to be maintained past the original author (i.e. survives uncertainty).
    \item Be able to respond to users' needs, even if they change over time (i.e. supports relevant concerns).
\end{enumerate}

As outlined in \secref{sec:stack}, even for \gls{software} that is maintained, for example by a third party organization (e.g. Harris Geospatial Solutions for \gls{IDL}) does not guarantee future usage of this technology within astronomy \citep{momcheva15}. As astronomy shifts towards a more community-developed, open source set of tools, it iss critical that different constituents of the astronomy community develop an understanding of the origin of this \gls{software} and how they might be able to participate in its development, maintenance, and long term sustainability:

\textbf{Software consumers (individual astronomers)}: Most individual researchers are \emph{consumers} of \gls{community software}, that is, they make heavy use of the \gls{software} tools developed by their peers but do not routinely participate in the development of the software. Like most community-developed open source projects, this is the norm and is acceptable. However, complete ignorance of the origin of the \gls{software} they are using creates a risk to the sustainability of the projects and individuals responsible for creating the software. For example, if they do not realize the \gls{software} they are using comes from other researchers, they may not support hiring, tenure, etc of those who build that \gls{software}, thereby stopping them from producing and maintaining the \gls{software} itself.  We believe therefore that even as \gls{software} consumers, astronomers should increase their awareness of the origin of the \gls{software} they are using and realize that they have an important role to play in the community by 1) providing feedback to \gls{software} projects by filing bug reports, feature requests, feedback on existing tools, and perhaps contribute other resources like documentation if they have relevant expertise; 2) recognizing that \gls{software} is created by \emph{people}, and that supporting the work of their peers  (be it financially, socially, or even emotionally) who spend time creating these tools is necessary for the tools they use to even exist; and 3) recognizing and advocating for the broader concept that using a shared set of community tools can improve all of science for less money.
\newline

\textbf{Individual \gls{software} creators (individual astronomers and engineers)}: While these are the bread-and-butter \emph{of} these community efforts, they are not without shared responsibility here.  Specifically, the builders of community have a responsibility for being aware of the community they are building for.  E.g. they need to remember that the user community typically does not have as much technical expertise and therefore requires their \emph{help} to both learn how to use the \gls{software} and understand why it is useful. They also need to understand the unique responsibility that creating \gls{software} sustainably is work (see the above subsections) and must either agree to such work or communicate clearly to their potential users that they cannot do it without help.

\textbf{Institutional \gls{software} creators (projects/missions/facilities)}: Observatories and missions (e.g. \gls{LSST}, \gls{JWST}, \gls{DKIST}), especially in development \& construction phases, spend significant resources developing \gls{software} both for internal operations but also for their community to analyze and interpret data products from their facilities. These \gls{software} creators need to be incentivized to \emph{upstream} (i.e. contribute back new innovations to \gls{community software} packages) their \gls{software} where possible, thereby contributing to the large ecosystem of \gls{software} available to the general astronomy community. As discussed earlier in \secref{ssec:forcem}, \gls{community software} can be a force-multiplier when done right, but in order for this to happen, \gls{software} projects must recognize their role in the \gls{community software} ecosystem and shift towards being active contributors rather than consumers/users of community software.

\section{
Analysis Methods: Algorithms and Statistical Foundations
\Contact{Brian Nord and Andrew Connolly}}
\Contributors{Brian Nord <\mail{nord@fnal.gov}>, Andrew Connolly <\mail{ajc@astro.washington.edu}>, Yusra AlSayyad, Jamie Kinney, Jeremy Kubica, Gautham Narayan,   Joshua Peek, Chad Schafer, Erik Tollerud }\label{sec:algorithms}

\bigskip
\noindent\textbf{Note:}
If you have come directly to this chapter we suggest you please read at least the Introduction in \secref{sec:intro} before delving further.
\newline

\subsection{Recommendations}
    \nrec{Analysis}{Agency, Astronomer}{Medium}{algmodels}{
    Create funding models and programs to support the development of advanced algorithms and statistical methods specifically targeted to the astronomy domain} {The increasingly large and complex datasets resulting from a new generation of telescopes, satellites, and experiments require the development of sophisticated and robust algorithms and methodologies.
    These techniques must have statistically rigorous underpinnings as well as being adaptable to changes in computer architectures.}
    \nrec{Analysis}{Technologist, Astronomer}{Long}{discengine}{
    Build automated discovery engines}
    {New hypotheses are difficult to generate in an era of large and complex datasets.
    Frameworks that can detect outliers or new patterns within our data could address many of the needs of current and planned science experiments.
    Funding and developing these engines as a community would lead to broad access to the tools needed for scientific exploration.}
    \nrec{Analysis}{Agency, Manager, Astronomer}{Long}{algconnect}{
    Promote interdisciplinary collaboration between institutions, fields, and industry}
    {Expertise across multiple domains are required to tailor algorithmic solutions to astronomical challenges.
    The astronomical community should more heavily and directly engage researchers from industry and non-astronomy fields in the development and optimization of algorithms and statistical methods.
    Agencies and academic departments should develop funded programs to specifically connect astronomers to these experts through sabbatical programs, centers, fellowships, and workshops for long-term cross-domain embedding of experts.
    }
    \nrec{Analysis}{Agency, Astronomer}{Medium} {algeducation}{
    Develop an open educational curriculum and principles for workforce training in both algorithms and statistics}
    {The speed of model and algorithm evolution requires regular training and education for scientists and for those seeking to enter science.
    Developing and maintaining open curricula and materials would enable the teaching of algorithms and methodologies throughout the astronomical community.}
    \nrec{Analysis}{Astronomer, Agency}{Short} {algpub}{
    Encourage, support, and require open publication and distribution of algorithms}
    {The rapid adoption of advanced methodologies and the promotion of reproducible science would be significantly enhanced if we mandated the open publication and distribution of algorithms alongside papers.}


\subsection{Overview}
The paradigms for data analysis, collaboration, and training have simultaneously reached a watershed moment in the context of algorithms and statistical methods.
The onset of large datasets as a scientific norm accentuates this shift, bringing both technical opportunities and challenges.
For example, the development of new algorithms and data modeling techniques has recently accelerated dramatically, providing new modalities for investigating large datasets.
As this corner has turned in algorithmic development, the incorporation of rigorous statistical paradigms must keep apace.
However, this shift has just begun, and we still lack the tools to even contend with, much less fully take advantage of, increasingly complex datasets for discovery.

The paradigm shifts also bring organizational challenges that highlight issues with cultural norms of education and collaboration about development of data analysis techniques.
Discovery often occurs at the intersections of or in the interstices between domains, and therefore
multi-dimensional collaboration has irrevocably become a key component of research.
We need improved collaboration paradigms to take advantage of this accelerating emergence of technologies, thereby increasing the permeability of the barrier between different areas of science, and between academia and industry.
Moreover, innovation in methods of education and training in new analysis techniques lag behind the development of the techniques themselves, leading to growing unequal distribution of knowledge.
Similarly, accompanying \gls{software} development strategies must keep apace with these developments, both to ensure results are robust and to make sure the education and training can be equitably distributed.

We have an opportunity to act as the changes set in and leverage our community's energy and inspiration to initiate change in how drive algorithmic discovery in the petabyte era.
There is an opportunity for astronomy to both benefit from and help drive new advances in the emerging technologies.
Below, we discuss the key challenge areas where we can and provide possible directions for what we can do.

\subsection{Discovery in the Petabyte Era}

At present the process of hypothesis generation in astronomy has two pathways.
One is theoretical, wherein predictions from theory provide hypotheses that can be tested with observations.
The other is observational, wherein surprising objects and trends are found serendipitously in data and later explored.
As theory comes to depend on larger and larger simulations, and observational datasets grow into and beyond the petabyte scale, both of these pathways are coming under threat.
With such large datasets, classical modes of exploration by a researcher are becoming prohibitively slow as a method to discover new patterns in data (e.g. finding objects and correlations by plotting up datasets).
Without new hypotheses (and ways to develop them) in the 2020s, there may be no astronomy in the 2030s.

A key example of the challenge lies in explorations of high-dimensional datasets.
Long ago, the discovery that stars fill an approximately \gls{1D} space in magnitude-color space led to a physical model of stellar structure.
This is a low-dimensional, non-linear representations of higher-dimensional data.
Indeed, seemingly smooth structures in astronomical data can have surprising substructure (e.g. the Jao/Gaia Gap \citep{Jao_2018}).
1D gaps in famous \gls{2D} spaces are visually discoverable.
However, we lack comparable methods to find \gls{2D} gaps in \gls{3D} spaces, let alone structures in the extremely high-dimensional data that modern surveys create.
Recently, \cite{2018AJ....156..219S} found 17 pure blackbody stars \emph{by eye} amongst the 798,593 spectra in \gls{SDSS}, nearly two decades after they were acquired.
This result shows both how interesting outliers can be, and how by-eye methods are slow and not practical at the petabyte scale.
With trillions of rows available in upcoming surveys, we'll have the ability to find low-dimensional substructure in high-dimensional that has potential to yield new physical insight --- but only if we have the tools to do so.


As an example of such a tool, purpose-built Machine Learning (\gls{ML}) algorithms coupled with deep sub-domain knowledge can successfully expose hitherto unknown objects that can significantly advance our understanding of our universe \citep[e.g.][]{2017MNRAS.465.4530B}.
Unfortunately, any successful exploration requires a) deep algorithmic and implementation knowledge b) deep physical and observational domain knowledge and c) luck.
Deep algorithmic knowledge is necessary as off-the-shelf algorithms usually need significant adaptation to work with heteroscedastic and censored astronomical data.
Deep observational domain knowledge is needed as outlier objects are often artifacts and surprising trends may be imprints of the data collection method.
Deep physical domain knowledge is needed to make sense of the result, and understand its place in the cosmos.
For example, algorithms to find low-dimensional structures \citep[e.g. Manifold Learning;]{2016arXiv160302763M} are only one piece.
Observational expertise is necessary to determine that the observed manifolds are real, and astrophysical expertise is necessary to formulate physical explanations for the observations.
Finally, not all searches will return results; a modicum of luck is needed.
This trifecta of algorithmic knowledge, domain knowledge, and luck is rare.


Over the next decade, we expect astronomy to require unique, fundamental new developments in algorithms, statistics, and machine learning.
Despite the incredible pace of innovation within these fields, it will not be enough for astronomy to ride along and adopt general technologies.
Astronomy’s science drivers will bring unique algorithmic and statistical questions, data characteristics, and edge cases that will both require and drive continued investment and innovation.

We argue that the path forward is through the construction of intuitive, trustworthy, robust, and deployable algorithms that are intentionally designed for the exploration of large, high-dimensional datasets in astronomy.
When we consider the current landscape of astronomical research and the upcoming generation of sky surveys, we can already identify areas where algorithmic and statistical investment are needed, such as:
\begin{enumerate}
\item Online (i.e. close to real-time) alerts and anomaly detection in large sky surveys will require high throughput algorithms and models in order to keep up with the volume of data produced.
\item Statistical and learned models need to go beyond black box optimization. Models should be understandable and interpretable in terms of the physical systems they represent.
\item Machine learning algorithms may need to be adapted to make effective use of domain knowledge such as physical constraints and data collection methodology.
\item Machine learning techniques often introduce new parameters that must be recorded in a standardized form to allow other researchers to reproduce analysis.
\end{enumerate}

Very few researchers have both all the needed skills and the bravery/foolhardiness to seek out risky avenues of research like these.
We therefore propose that funding agencies fund the creation and maintenance of ``discovery engines'' --- tools that allow astronomers without deep algorithmic knowledge to explore the edges of data spaces to hunt for outliers and new trends.
These engines should be hosted near the data when needed (\secref{sec:coloc}), but should be initiated by the astronomical and methods-development communities.

The development of new statistics and algorithms can be accomplished through a variety of methods, including: on-boarding dedicated algorithmic/data-intensive science experts onto astronomy teams, facilitating partnerships (with industry or other academic fields), and building internal expertise within the community through education and training.
Regardless of the mechanism, it is important that the development of new statistical and algorithmic techniques is considered a core part of astronomical missions.

\subsection{The state of statistics: statistical methodologies for astrophysics
}

Statistical methods and principles are the backbone upon which successful estimation, discovery, and classification tasks are constructed.
The tools commonly associated with Machine Learning (e.g. deep learning) are typically efficient, ``ready-to-use" algorithms (albeit with ample tuning parameters).
On the other hand, statistical approaches employ a set of data analysis principles.
For example, Bayesian and frequentist inference are two competing philosophical approaches to parameter estimation, but neither prescribes the use of a particular algorithm.
Instead, the value (and perhaps the curse) of the statistical approach is that methodological choices can be tailored to the nuances and complexities of the problem at hand.
Hence, when considering the statistical tools that are crucial for astronomy in the coming decade, one must think of the recurring challenges that are faced in data analysis tasks in this field.

Further, as the sizes of astronomical survey datasets grow, it is not sufficient to merely “scale up” previously-utilized statistical analysis methods.
More precisely, modern data are not only greater in volume, but are richer in type and resolution.
As the size and richness of datasets increase, new scientific opportunities arise for modeling known phenomena in greater detail, and for discovering new (often rare) phenomena.
But these larger datasets present challenges that go beyond greater computational demands: they are often of a different character due to the growing richness, which necessitates new analysis methods and therefore different statistical approaches.
Hence, as the complexity of astronomical data analysis challenges grow, it is imperative that there be increasing involvement from experts in the application of statistical approaches.

To ground these ideas, in the following subsections we will consider examples of technical and organizational challenges that, if advanced over the next decade, would provide the greatest scientific benefit to astronomy.

\subsubsection{Technical Challenges}

\begin{enumerate}
\item {\it Methods for the analysis of noisy, irregularly-spaced time series.}
Future time domain surveys, like \gls{LSST},  will generate a massive number of light curves (time series) with irregular observational patterns and in multiple bands.
This goes beyond the limits of classic time series models, which assume regularly spaced observations with a simple error structure.
Areas of need include feature selection for classification, periodicity detection, and autoregressive modeling,
\item {\it Likelihood-free approaches to inference.}
Likelihood-based inference is standard in astronomy, but as the sizes of datasets grows, any flaw in the assumed likelihood function will result in a bias in the resulting inference.
Such flaws result from unwarranted Gaussianity assumptions, difficult-to-model observational effects, and oversimplified assumptions regarding measurement errors.
Likelihood-free approaches, such as approximate Bayesian computation, hold promise in astronomy, but much work is required to develop tools and optimize them for astronomy datasets and therefore make this computationally-intensive approach feasible.
\item {\it Efficient methods of posterior approximation.}
Even in cases where a likelihood function is available, constructing the Bayesian posterior is challenging in complex cosmological parameter estimation problems, because future inference problems will push the computational boundaries of current \gls{MCMC} samplers.
Work is needed to improve the performance of chains, which must adjust to degeneracies between cosmological parameters, handle a large number of nuisance parameters, and adhere to complex hierarchical structure that is increasingly utilized in such analyses.
\item {\it Emulators for complex simulation models.}
It is increasingly the case that a simulation model provides the best understanding of the relationship between unknown parameters of interest and the observable data.
Unfortunately, these simulation models are often of sufficient complexity that a limited number of simulation runs can be performed; the output for additional input parameter values must be approximated using emulators that interpolate these available runs.
Emulation to sufficient accuracy requires careful selection of both the input parameters for the training sample and the method of interpolation; both of these must be done with consideration of the particular application.
\item {\it Accurate quantification of uncertainty.}
Complex inference problems in astronomy are often, out of necessity, divided into a sequence of component steps.
For example, classification of Type Ia supernovae, a challenging problem on its own, is just a step in a larger analysis that seeks to constrain cosmological parameters.
Separately, redshifts and luminosity functions are estimated and then fed into larger estimation problems.
This divide-and-conquer approach requires careful consideration of the propagation of error through the steps.
How does one quantify errors in redshift estimates in such a way that these uncertainties are accurately accounted for in the downstream analyses?
How is contamination that results from misclassification of supernovae reflected in the uncertainties in cosmological parameters estimated from these samples?
LSST faces challenges of separating identifying images in which overlapping objects are ``blended"; how is the uncertainty inherent in this problem incorporated into analyses that use these images?
Careful consideration of such questions is crucial for attaching accurate statements of uncertainty to final estimates.
\end{enumerate}

\subsubsection{Organizational Challenges}

\begin{enumerate}
\item {\it Accessible publishing of methods.}
Advances in statistical theory and methods abound in the literature of that field, but it is often presented in a highly formalized mathematical manner, which obscures the aspects of most importance to potential users.
This creates a barrier to the appropriate use of these methods in astronomy.
The greater involvement of data scientists in collaborations will help to bridge this divide, and enable these individuals to make significant contributions.
This will require appropriate professional recognition for this effort, including encouraging the publication of methodology papers in astronomical journals by data scientists (see \secref{sec:iwft:cookies} for related workforce issues).
\item {\it Avoiding the ``algorithm trap."}
Astronomical inference problems are of sufficient complexity that full use of the data requires analysis methods to be adapted and tailored to the specific problem.
For this reason, statisticians prefer to not think of an analysis as the application of a ready-made ``algorithm.''
By contrast, astronomers are generally more interested in the result of the analysis, so are attracted to well-separated ``algorithms'' they can apply to a problem.
This difference in perspective only increases the need to have data scientists deeply involved in the collaborative process.
\item {\it Reducing barriers for statisticians.}
From the other side, data scientists face challenges in applying analysis techniques astronomical data.
This is partly due to technical difficulties like unique file formats and data access issues.
But it is also because deeply understanding the science is frequently crucial to building methods tailored to the problem, as outlined above.
More effort needs to be placed on reducing these barriers.
For example, astronomers can work to isolate important statistical aspects of larger problems and create user-friendly descriptions and datasets to allow statisticians to more quickly learn and focus on making a contribution.
At the same time, embedding statisticians and data scientists close to astronomers will help bring the former to a better understanding of the astronomy perspective.
\end{enumerate}

\subsection{The state of algorithms: developments in the last decade}

A key development that has enabled science in this past decade has been the development of a number of general purpose algorithms that can be applied to a variety of problems.
These algorithms, irrespective of what programming language the implementation is in, have made astrophysical research more repeatable and reproducible, and less dependent on human tuning.

For example, \gls{PSF} kernel convolution has enabled time-domain astrophysics, and is a key component of difference imaging pipelines, but is also used to generate deep stacks of the static sky, allowing us to find ever more distant galaxies.
These developments in turn have spurred the development of new algorithms.
In roughly 20 years, the field has moved from Phillip Massey's guide to doing aperture photometry by hand with \gls{IRAF} for small, classically scheduled programs, to completely automated surveys that optimize their observing schedule in real-time, record data, detrend the observations, and perform automated \gls{PSF} photometry of billions of deblended sources.

As with statistics, the distinction between algorithms, and the \gls{software} implementation of algorithms is blurry within the community. In many situations, we now use algorithms without any knowledge of how they work
For example, we can now expect to sort tables with millions of rows on multiple keys, without knowing the details of sorting algorithms, precisely because these details have been abstracted away.
We note that the many widely used algorithms, such as affine-invariant Markov Chain Monte Carlo techniques are widely used precisely because the algorithm is implemented as a convenient \gls{software} package.
Community-developed \gls{software} packages such as \texttt{scikit-learn}, \texttt{astropy}, and the \gls{IDL} Astronomy Library have increased the community's exposure to various algorithms, and the documentation of these packages has in many cases supplanted implementation-oriented resources such as Numerical Recipes.

At the same time in the broader world, a class of algorithms is being used to execute tasks for which an explicit statistical forward model is too complex to develop, and correlations within the data itself is used to generate actionable predictions.
These \gls{AI} techniques include machine learning models, which have been used to replace humans for tasks as varied as identifying artifacts in difference images, to categorizing proposals for time allocation committees.
These \gls{AI} techniques, in particular deep learning methods, are increasingly viewed as a solution to specific petabyte scale problems, as they have been successfully deployed in the commercial sector on these scales.
We anticipate increasing adoption of these algorithms, as user friendly implementations such as \texttt{pyTorch} and \texttt{Keras} become more well known, and data volumes grow.
It is also likely that the algorithms that are used to train these machine learning methods, including techniques like stochastic gradient descent, will find more use within the astronomical community.

Machine learning algorithms are necessary but not sufficient to continue the progress in astrophysical research that is driven by algorithms.
In particular, machine learning methods are often not-interpretable, and while their output can be used effectively, those outputs are not true probabilities. The scientific method fundamentally involves the generation of a testable hypothesis that can be evaluated given data, and is therefore inherently statistical. As data volumes grow, the dimensionality of models grows, and there is increasing recognition that the model structure is hierarchical or multi-level. While we see increasing adoption of hierarchical models for Bayesian inference, there remains much to do to increase awareness of algorithms to effectively evaluate these models, including probabilistic programming - algorithms that are used to build and evaluate statistical models in a programmatic manner.


As in the previous section, we now separately consider some of the specific technical and organizational challenges in the area of algorithms.

\subsubsection{Technical challenges}

\begin{enumerate}
\item Both algorithms and models need to be trustworthy and interpretable. It's easy to throw a dataset into a neural net or ensemble classifier and overfit.
Tools need to be developed that recognize these traps and in large-scale datasets, and bring them to the attention of the user.
\item Many algorithms, especially in the machine learning space, require labeled data that may not be available at sufficient volumes, or at all.
\item The \gls{reproducibility} of results derived from algorithms needs to be improved. This is especially important with machine learning models where black-box optimization is often used because it is an easy-to-provide feature. Such \gls{reproducibility} improvements could be as simple as defining standardized formats for how we document the model learning parameters, but could also be more complex, including building out tools that are designed specifically for \gls{reproducibility} (e.g. Data reduction pipelines with built-in provenance, or Jupyter notebooks that download their own data).
\item Scalability of newly-developed algorithms. With the data volumes of the petabyte era, efficiency in all parts of the stack is necessary.  Such optimizations are usually possible, but require investment of time (often by different people than those who develop the first iterations of the algorithm).
\item Astronomy data has some differences that can expand current algorithmic development at large. This particularly includes use of measurement uncertainties, as general-use algorithms often make assumptions that work for other fields that are homoscedastic or Gaussian which fail in Astronomy. There is also a need for more algorithms that account  for posteriors, a particularly strong need in astronomy because its domain of ``the universe as a whole'' means that algorithms applied to one dataset need their outputs to be considered by another.
\item Significant work is still needed in adapting and improving the current space of existing algorithms: optimizing traditional astronomy algorithms, adapting them for a \gls{cloud} setting, or even making small accuracy improvements.
\end{enumerate}

\subsubsection{Organizational challenges}

\begin{enumerate}
  \item It is difficult to get the necessary expertise onto all missions that will need it both in terms of developing the expertise internally (due to the fast pace of change in the space) and hiring in experts.
  \item There is currently no established marketplace/mechanism for matching difficult problems in the astronomy domain to relevant experts outside an astronomer's network.  This is particularly acute given the discussion above about the growing importance of statistical and data science expertise.
  \item There is a missing component in the conduit of moving new algorithms developed in academia into robust, usable, finished products. See \secref{sec:software} for additional discussion in this area.
  \item We need standardized processes for publishing algorithms and machine learning models such that the results obtained with these algorithms/models are: broadly accessible, discoverable, fully reproducible (including archiving the model parameters), and easily comparable with other algorithms in the problem space.
  \item We need to define and fund a process for continually modernizing/upgrading algorithms as the broader environment changes (new languages, new libraries, new computational architectures, shift to \gls{cloud} computing, etc). See \secref{ssec:sw_alive} for a broader discussion of mechanisms and recommendations for this.
\end{enumerate}

\subsection{Emerging trends in industry and other fields
}
\label{sec:algos:industrytrends}

Over the past two decades, the wider industry has also seen a shift in development approaches and computational techniques that can be adopted by the astronomical community.
As noted in \secref{sec:software} open source \gls{software} has become a new normal with communities sharing their investment in \gls{software} development.
When considered along with the industry's shift toward \gls{cloud} computing and \gls{software} as a service, astronomy can benefit from the new scale and availability of off-the-shelf solutions for computation and storage.
Astronomers no longer need to focus significant portions of time on the low-level technical details in running dedicated banks of computers to support each survey.

This service model is being extended beyond \gls{software} deployments and starting to push into algorithms as a service.
Cloud machine learning services provide a portfolio of general algorithms.
Instead of worrying about the specifics of the algorithm development, users focus only on model specification.
This requires a shift in how we think about new algorithm development.
Instead of focusing on the details such as implementation, optimization, and numerical accuracy, the practitioner focuses primarily on the high level model specification.
Due to a series of recent successes, a significant focus within hosted machine learning services has been on deep neural networks (DNNs).
NNs have shown remarkable success across a variety of tasks.
Further new developments such as convolutional neural networks and recurrent neural networks have extended the power of this technique.

Another area of focus within the field of machine learning is blackbox optimization.
Techniques such as Gaussian decision processes, allow algorithms to jointly model and optimize unknown functions.
These techniques can be applied to a range of problems from optimizing real-world, physical processes to optimizing the parameters of a machine learning system (e.g. AutoML).

The ultimate goal of algorithms as a service can be seen in the advancements in AutoML.
AutoML systems aim to abstract away not just the algorithm’s implementation details, but also the need to manually tune model parameters.
For example, recent work in \gls{NAS}, allows the AutoML system to handle such development decisions as choosing the structure of the network (number and width of layers) as well as the learning parameters.
While this automation greatly simplifies the problem of constructing accurate models, it does move the practitioner one step further from understanding the full details of the model.

There is an opportunity for astronomy to both benefit from and help drive new advances in the emerging industries.
As noted above, astronomy can benefit from the shift from individually developed and maintained systems to hosted platforms that allow more effort to be spent on the data analysis itself.
Moreover, the shape and size of science data serve as a driver for the development of new algorithms and approaches.
We expect many of the upcoming advancements to be driven by real-world problems—machine learning will rise to the challenge of solving new, open problems.
The recommendations in this chapter aim to ensure some of these problems and solutions are in the astornomy domain.


\subsection{Enhancing Interdisciplinary Programs and Collaborations}

The past decade has been a period of rapid change in the the multi-dimensional landscape of algorithms, computing, and statistics. We have seen the rise of new ``standard'' programming languages and libraries (e.g. Python, \texttt{astropy}, \texttt{scikit-learn}).
There has been a proliferation of new algorithmic and statistical techniques --- from improvements in image processing and compression to the rise of deep neural networks as a powerful tool from machine learning.
We have seen the rise of new computational modalities, such as \gls{cloud} computing and \gls{software} as a service.
New distributed compute frameworks such as Dask and Spark are emerging to process and analyze large and complex datasets.
Even the basic mechanics of computation is undergoing a shift with the availability of specialized hardware such as GPUs and TPUs, requiring a new domain of knowledge to efficiently deploy solutions.
There is no reason to expect the pace of innovation to drop off anytime soon.

This rapid pace of advancement means that it is no longer possible for a single astronomer or even a small team of astronomers to build the necessary depth of expertise in all of these areas.
However, these technologies are already proving critical for maximizing the scientific reach of new research.
Robust methodologies that can scale to the expected size and complexity of the data from new astronomical surveys and experiments will need to be accessible and usable by a broad section of our community.
As new technologies spring up quickly,  the astronomical community will need to balance the cost of learning the new technologies with the benefits they provide.
It is not reasonable to expect every astronomer to keep up with all of the advances.
A number of new ad hoc collaborations or collectives have sprung up to bring together astrophysicists and deep learning experts, such as the Deep Skies Lab\footnote{\url{deepskieslab.ai}} and Dark Machines\footnote{\url{darkmachines.org}}.

In cases where collaborations exist today, there can be a variety of complicating challenges.
There is currently no established marketplace for matching difficult problems in the astronomy domain to relevant experts outside an astronomer's network (see also \S\ref{sec:algos:industrytrends}).
The resulting in-depth collaborations have start up overhead as the external experts learn enough about the problem domain to be helpful. Short-term engagements can suffer from a lack of depth or insufficiently productionized solutions.
Even in longer term engagements, there can be misalignment between the parties due to the different incentives.
For example, statisticians and computer scientists in academia are primarily recognized for only the novel contributions to their own fields.
Papers that apply existing methodologies to new problems are not considered significant contributions to their fields.
Similarly, members of the astronomy community are not fully recognized for their algorithmic contributions.

There are many opportunities for astrophysics to benefit from these investments in technology and computational algorithms.
However, requires that we change how astronomy engages with experts in other fields.
The exact shape of this engagement can take a variety of forms. Examples include:
\begin{enumerate}
\item Provide funding for astronomical missions to engage with external experts (academic or industrial) via consulting, co-funded research, or subcontracting.
\item Encourage a robust community of volunteers via open source contributions and engagement.
\item Create forums for external methodological experts to engage in astronomical projects and analyses.
Data challenges and hack sessions can be used to encourage engagement, but they require sufficient organization and communication (i.e. funded effort) to ensure they can engage \gls{software} engineers at an appropriate level.
\item Encourage recognition of interdisciplinary contributions within academic areas (e.g. career progression for statisticians that enable new astronomy without necessarily creating new statistics).
\item Organize workshops that bring together members of these different fields and can facilitate matching along problem domain.
\item Provide funding for astronomical programs to hire full time experts to be embedded within the mission.
It is important to note that this approach comes with challenges in recruiting (both these areas are in high demand), costs of attracting high quality personnel, and in stability for the team members (the algorithmic / statistical workload might not be consistent throughout the life of a project).
\item Implement reverse sabbaticals where experts from  industry can embed in projects for short intervals (a few months).
\item Train astronomers in these fields to become resident experts.
Encourage mobility of these experts to provide support for new missions.
\item Establish a center for algorithmic and statistical development in astronomy (centralized or virtual) that employs full time experts in fields such as algorithms, statistics, and machine learning. This center would be a community resource that provides support to individual programs via deep engagement.
\end{enumerate}

The goals of these interactions are not to provide programming support for projects but to develop a base of expertise built from academic and industrial experts that can help to define, design, and guide the development of computational and statistical projects within astronomy.
The form and depth of the engagement will naturally be project dependent.
Experimental and privately-funded interdisciplinary centers e.g.\ the Moore-Sloan Data Science Environments at Berkeley, \gls{NYU} and the University of Washington, or the Simon's Flatiron Institute have demonstrated how expertise in data science can advance a broad range of scientific fields.
Access to the resources at these centers is, however, limited to researchers at these privileged institutions.
The challenge we face is how to scale these approaches to benefit our community as a whole.

\subsection{Education and training}

Training a workforce that can address the algorithmic and statistical challenges described in this Chapter will require a significant change in how we educate and train everyone in our field, from undergraduate students to \gls{PI}'s.
The discussion in this section is complementary to and aligned with that found in \secref{sec:iwft} and \secref{sec:epo}.
The traditional curricula of physics and astronomy departments do not map easily to the skills and methodologies that are required for complex and/or data intensive datasets.
This is a rapidly changing field, and will remain so for at least a decade.
However, a strong foundation in Bayesian statistics, data structures, sampling methodologies, and \gls{software} design principles would enable professional astronomers to take advantage to big data in the next decade.
Bridging this gap between the skills we provide our workforce today and the ones they might need to succeed in the next decade should be a priority for the field.

In the previous decade there was substantial progress in creating material to support the teaching of statistics and machine learning in astronomy.
This includes the publication of introductory textbooks \citep{astroMLText,R:Kohl:2015en,rfaq}, the creation of common \gls{software} tools and environments \citep{astropy1}, the development of tutorials, and a growing focus on \gls{software} documentation \citep{astropy_tutorials}.
The emergence of Jupyter \citep{jupyter} as a platform for publishing interactive tutorials and Github and Gitlab for hosting these tutorials and associated code has simplified the process of sharing material.
To date, however, there has been little coordination in this effort.
The coverage of topics in the available material is not uniform.
Moreover, the underlying principles and foundations of statistics are often not covered in favor of the introduction of commonly used \gls{software} tools and algorithms.
For the case of algorithmic design and optimization there has been substantially less progress in training the community.
Instead, the primary focus being the development of introductory materials such as the Software and Data carpentry \citep{data_carpentry, 2013arXiv1307.5448W}.

We have started to make progress in providing an educational foundation in statistics and algorithms, but it is not uniformly available across our community --- with significantly less access at smaller colleges and in underrepresented communities.
We, therefore, recommend the development and support of a common and open set of educational resources that can be used in teaching statistics, and algorithms, and machine or computational learning. Determining what constitutes an appropriate curriculum will be a balance between providing the foundations of statistics and algorithmic design appropriate for the broader science community and teaching specialized skills (e.g. optimization, compilers) that may benefit a smaller, but crucial, set of researchers who will engage in the development and implementation of computing and \gls{software} frameworks.

This will likely require a coordinated effort to integrate current resources within a broader curriculum and to make them easily accessible --- in a manner where anyone, from astronomer to an entire educational institution, can create custom courses tailored to their needs.
Given the rapid evolution in algorithms and in the ecosystem of tools over the last decade, and looking to the future, this curriculum will need to be able to evolve.

\section{Workforce \& Career Development \Contact{Dara Norman}}
\Contributors{Dara Norman <\mail{dnorman@noao.edu}>, Kelle Cruz, Vandana Desai, Britt Lundgren, Eric Bellm, Frossie Economou, Arfon Smith, Amanda Bauer, Brian Nord, Chad Schafer, Gautham Narayan, Ting Li, Erik Tollerud}
\label{sec:iwft}

\bigskip
\noindent\textbf{Note:}
If you have come directly to this chapter we suggest you please read at least the Introduction in \secref{sec:intro} before delving further.
\newline

\subsection{The growing importance of a tech-savvy workforce of astronomers}
In the rapidly approaching era of large surveys, experiments, and datasets, we will only reach our scientific goals if we train and retain highly capable scientists, who are also engaged with technological advances in computing.
With the goal of advancing scientific discovery through the collection and analysis of data, we must commit and dedicate resources to building both the skills and competencies of this workforce.
 We define the workforce as those who derive science from the data, as well as those who build the infrastructure to collect and prepare the data for discovery analyses; one cannot happen without the other.
The areas and skill sets in which our teams need training are \gls{software} carpentry, algorithms, statistics, the use of software tools and user services; and \gls{software} engineering effective practices, data management and access (for support staff).
In addition, all members of the workforce, including those who provide infrastructure, will require professional development opportunities and support for their career advancement allows them to remain as productive and respected members of the field.

In this chapter we discuss the activities needed to build, support, and advance the scientific workforce that will take the petabytes of data collected to scientific discoveries over the next decade. In particular, \secref{sec:iwft:demographics} discusses the current demographics of the data science support mission while (\secref{sec:iwft:software}) exemplifies the scope of training that is needed to build this workforce. \secref{sec:iwft:datascience} focuses on training for researchers who are more accurately described as ``users.''  In \secref{sec:iwft:careers}, we discuss modern challenges for these career paths, as well as how to address them. Finally, in \secref{sec:iwft:credit}, we identify  metrics that we should be using for training in career development and for reviews in career advancement.

\subsection{Demographics - who is this workforce and where are they now\label{sec:iwft:demographics}}
Data support roles are spread throughout the astronomy and astrophysics (hereafter, ``astronomy'') scientific community, and encompass people with a variety of job types and descriptions at levels from post-baccalaureate to PhD.
A range of experience with either topics of astronomy or computing also differentiate roles.
This range of data support positions requires a diversity of opportunities for training to work at the various levels. In addition, different career development and advancement suited to those career tracks is also required.
For example, positions for those with PhDs are significantly different from those that require only a post-bac degree, and thus the metrics used to support and determine career advancement must also be different.  It has only recently been recognized that this role should be trained for and tracked independently of scientific interests and other professional duties.
Consequently, the community has not adequately tracked the quantity and demographics of astronomy researchers currently engaged in science data support roles.

Instead of quoting statistics here, we present exemplar descriptions of current job titles and roles.
Many of the people engaged in science data support hold PhDs in astronomy, astrophysics or physics.  These researchers may be employed at colleges, universities, data centers or observatories, national laboratories. They may hold a leveled academic title (e.g. Professor, Astronomer, Scientist, etc.), as well as an additional functional job position in centers or programs with names ``Data Science Mission Office,'' ``Community Science and Data Center,'' ``Infrared Processing and Analysis Center.''
 Despite the distinct nature of the work for scientists who hold these multiple roles/titles, the same metrics (e.g. numbers of published papers, h-value, etc.) for career advancement through titled positions (i.e. assistant, associate, full, etc.) are used for them as for faculty with more typical duties and responsibilities.
More discussion is in \secref{sec:iwft:careers}.

There are also many other science data support roles, in which staff have degrees at the BS, MS, or PhD level with position titles like ``research and instrument associate,'' ``research and instrument scientist,'' ``mission systems scientist,'' ``archive scientist.'' These staff are often responsible for coding and database support.
Below, we discuss the resources and cultural changes needed to support the career trajectories of this workforce, to slow the threat of ``brain drain'' from the field, and to develop a workforce that can thrive in academia, industry, or government lab positions.

\subsection{Training to contribute to \gls{software} development: Building the next generation\label{sec:iwft:software}}

Astronomers have a long history of developing useful \gls{software}, but \gls{software} development itself has not been considered a core component of the astronomy training curriculum. The expectation of petascale datasets in the 2020's provides a strong motivation to increase familiarity with effective practices in \gls{software} development, as well as with existing frameworks that are widely used in the commercial sector. This cultural change will lead to better \gls{software} in astronomy and more innovative scientific discovery. It will also provide astronomers with invaluable training that will increase their familiarity with (and marketability to) work in industry.

Currently, effective practices include using \gls{version control} (e.g. GitHub), maintaining documentation and unit tests with code, and employing continuous integration methodologies, in which code is built and executed in shared repositories, allowing teams to identify issues early. Analysis in the 2020s will involve many pieces of \gls{software} that are integrated into complex pipelines, processing ever-larger volumes of data. Astronomical projects are now comparable in scale to large industrial \gls{software} development projects. Consequently, the gap between these effective practices and the modern cultural norm in astronomy and astrophysics must be reduced as the field transitions to increasingly large collaborations.

The increasingly critical role of \gls{software} development in astronomy clearly indicates it is crucial that \gls{software} development become part of the core graduate curriculum alongside typical coursework, like mathematics and observing techniques. Such coursework will also help reduce the disparity between students from diverse backgrounds, some of whom may never have been exposed to \gls{software} development, or even coding, as undergraduates. This course material complements, but is distinct from, training in data science and scientific computing techniques, which are increasingly being incorporated into Astronomy coursework. Developing the course material for data science work is likely beyond the scope of most departments, but vital steps have already been taken by several groups. Notably, the \gls{LSST} Data Science Fellowship Program has already developed materials to expose students to best practices for \gls{software} development. Curating these materials, and augmenting them with information for distribution on widely-used platforms will reduce the barrier to adopting such coursework or integrating it into existing classes.

Besides developing course material, there are several other challenges to supporting scientific \gls{software} training in a university setting. Another challenge is the lack of access to state-of-the-art technologies. Because the landscape of coding and \gls{software} development changes rapidly as coding languages come and go, workflow best practices continually evolve, and new platforms emerge and gain wide acceptance. For principal investigators and project managers to make informed decisions and guide their teams, there must be opportunities for them to stay abreast of these developments and to evaluate their utility even if they are not the ones actually using the various tools.

Yet another challenge resides in the structure and processes of university departments. Many computer science departments do not teach the programming skills necessary for scientists.
Current  faculty, especially at smaller, under-resourced schools, do not have the resources to develop these.
Thus, the burden of developing more appropriate materials is fractured and currently falls upon individual instructors. The field needs dedicated staffing to develop curriculum materials for computational training. A fundamental barrier to the development of reliable, curated, and widely shared \gls{software} in astronomy is the lack of incentives for this work and the dominance of the ``publish or perish'' mentality. Changing this cultural norm requires that our community incentivize --- both within scientific projects and across the field at the employment level --- work in developing good \gls{software} and in educating people to build good \gls{software}.
Two key updates must be made to our community's value structure to change cultural norms and prepare for the software challenges of projects in the 2020's: 1) software development work should be a part of assessments in service and research; 2) software that is widely used for scientific work should be well-written and -documented.
A full solution cannot be realized through universities alone, and partnerships with data centers, observatories, national labs, and professional societies are crucial.

The clear successes and popularity of the existing training programs, which grew organically out of the community, attest to the need for additional and more advanced training resources. While there are several successful programs that address some of these concerns, they are insufficient to meet the needs of the larger community.  For example, the Software Carpentry curriculum\footnote{\url{https://software-carpentry.org/lessons/}} is limited to the very basics of \gls{version control} and collaborative \gls{software} development but does not cover topics, like performance optimization, continuous integration and testing, and documentation. Furthermore, most of these workshops are targeted to senior graduate students, with a few targeting very early-career scientists, and they are not designed to meet the needs or concerns of mid-career scientists and managers. Thus, these programs are currently limited to a very small portion of the community and are currently unable to  provide the needed training to people in multiple sectors of our community who need and want these opportunities.

Staff at Data Centers may themselves currently lack up-to-date data science skills and  knowledge. Funding to support career development for current staff and to provide resources for centers to hire staff that have data science expertise is critical to building workforce capacity in the 2020s.

Fundamental coding and \gls{software} development skills are becoming increasingly necessary for success in every aspect of Astronomy. However, acquiring professional training in these skills is rare and inaccessible or impractical for many members of our community. Students and professionals alike have been expected to learn these skills on their own, outside of their formal classroom curriculum or work duties. Despite the recognized importance of these skills, there is little opportunity to learn and build them --- even for interested researchers. 
To have a workforce capable of taking advantage of the computational resources and data coming in the next decade, we must find and support ways to make coding and \gls{software} development training widely accessible to community members at all levels.

 \nrec{Workforce}{Agency }{Long} {wc3}{Programs to cultivate the next generation}{Agencies should fund more and large-scale programs that cultivate the next generation of researchers versed in both astrophysics and data science, similar to smaller and over-subscribed programs like Software and Data Carpentry, \gls{LSSTC} Data Science Fellowship/ La Serena Data School for Science, Penn State Summer School in Statistics for Astronomers.}

 \nrec{Workforce}{Agency }{Short} {wc4}{Support to produce training materials}{Provide funding to data and computational centers to produce modular and re-usable training resources to the community. These resources should be designed to be used by individuals, integrated into formal classes, and used as part of professional development training.}

 \nrec{Workforce}{Agency }{Long} {wc5}{Long-term curation of materials}{Funding must be provided to host and support educational materials in a long-term, stable, scalable place. Provides stability and improves discoverability if materials can live in a centralized location.}

 \nrec{Workforce}{Agency }{Medium} {wcfundpartner}{Funding for innovative partnerships}{Incentives should be provided to launch opportunities to harness partnerships between data centers, universities and industry through funding. For example, support for sabbatical programs at the data centers where teaching faculty can learn skills, develop educational materials for community use, and bring back to their home institutions.}

 \nrec{Workforce}{Astronomer, Educator, University }{Medium} {wc6}{Software training as part of science curriculum}{Individuals, departments, and professional societies should encourage educational programs to incorporate \gls{software} training skills into their existing courses and programs.}





\subsection{Training to take advantage of big data for research: Bytes to Science\label{sec:iwft:datascience}}
Astronomers who came of age before the era of Big Data require training to take advantage of astronomical “Big Data” in the 2020s. They also need these skills to mentor students, who are simultaneously learning both astrophysics and the uses of data for research. It is crucial that access to this training be made widely available to professionals who come from a variety of science backgrounds and are based at a broad range of institutions (e.g. universities, data centers, etc.). This is especially important, considering these professionals will be cultivating their students and the next generation of scientists, as well as making decisions about which technologies to invest in. If access to advancing data skills remains difficult to obtain, we will fail to build a diverse workforce equipped to answer the most pressing questions in astronomical research. Data Centers could play an important role in providing this training.

New, freely accessible open source code and Jupyter frameworks like SciServer.org and \gls{NOAO} Data Lab enable anyone with a web browser to quickly and easily analyze vast stores of professional astronomy datasets via web-based notebooks.  These cloud-based platforms can democratize educational access by providing a scale of computing power and data storage that was previously reserved for students and faculty at well-resourced research institutions, where high-performance computing access and support are abundant.  A small number of astronomers in higher education are already developing instructional activities for these platforms. These instructional materials train students and other users to explore and analyze large professional astronomy datasets with ease and to equip users with the computational foundation needed to pursue advanced independent research projects.

Jupyter notebooks in particular hold enormous potential for training the current and next generation of astronomy professionals. However, currently, the development of standardized curricular activities is performed in an entirely ad-hoc manner.
Limited resources (funding and time) lead to very little deliberate coordination amongst various astronomy faculty who produce such materials, and these products are not sufficiently discoverable (e.g. accessible through a common repository).


The establishment of Community Science Centers hosted by Data Centers (like \gls{NOAO})  can be a hub (clearing house) to bring information to the community about opportunities for the kind of resources and training that allow a broad group of researchers to go from petabytes to publications.




In order to provide the most useful training, data centers need a clear view of user needs. This information is provided by advisory committees, like ``User Panels.'' However, these panels are traditionally populated by astronomers based at R1 institutions and other data centers. Data Centers should ensure that their User Panels include representatives from small and under-resourced institutions; this will provide a clearer picture of the unique training needs and challenges that must be addressed for these researchers.  In addition community surveys that reach astronomers who do not currently use data centers should be undertaken to better understand what barriers exist.

 \nrec{Workforce}{Agency }{Short} {wc1}{Training activities and materials}{Agencies must ensure that big data science training activities and materials for PROFESSIONALS (as well as students) are included as part of the federally funded data center’s mission and deliverables.}
 \nrec{Workforce}{Agency }{Medium} {wc2}{Change advisory board representation}{Federally (and privately?) funded science centers should include representatives from small and under-resourced institutions to provide a broad and clear picture of need in the community.  The collection of information, perhaps through surveys, to better understand the barriers to access that exist for astronomers at these institutions should be undertaken by data centers and others.}

\subsection{Identifying career paths around scientific \gls{software} development \& big data science\label{sec:iwft:careers}}
The key skills necessary for data-intensive scientific research are also highly valued in industry, government, and media/communication sectors. Astronomy training can serve as a stepping stone to fulfilling careers in a wide variety of fields, and astronomers should support and encourage those who transition to jobs in non-academic science, because ties with industry can strengthen and leverage our partnership opportunities. However, we need informed people on both sides: in many cases, challenging and uncertain career paths in astronomy push the best and brightest towards careers where their contributions are more readily appreciated. This ``brain drain'' siphons away the very researchers most needed to tackle the most pressing science questions of the 2020s.

\subsubsection{Universities}
In the university context, tenure-track faculty positions remain the gold standard for stability, compensation, and prestige.  However, despite the  fundamental role of \gls{software} in scientific discovery, it remains difficult to receive credit towards tenure and promotion for developing \gls{software} and services.  \secref{sec:iwft:credit} offers more specific recommendations for improving recognition for these contributions.

Even with appropriate credit for \gls{software} contributions, faculty positions will continue to carry expectations of leadership, grant-writing, teaching, mentorship, and service, as is appropriate.  Furthermore, driven by ongoing changes in the landscape of higher education, tenure-track hiring continues to flatten.  To benefit from the opportunities of large datasets, universities also need the ability to support and retain technically capable faculty and staff, who have expertise and a longevity that typically cannot be matched by graduate students or postdocs.  These “Research Software Engineers” (\cite{rse}) would provide a technical core for data-intensive research groups, just as opto-mechanical and electrical engineers are vital to the success of instrumentation labs.

Stable funding is the largest need for the success of staff Research Software Engineers \citep{geiger_2018}. A patchwork of 2-3-year soft-money grants is insufficient to retain highly-capable professionals, especially when industry salaries are significantly higher. Universities should explore means of providing internal support for data science staff, perhaps sharing capacity between academic groups or departments.
Long-term vision and leadership in the field are needed to recognize and measure relevant metrics and make them part of advancement/career ladders.

 \nrec{Workforce}{Manager }{Short} {wcA}{Recognize  software as part of the career path}{Software should be recognized in hiring and career development as a core product of modern astronomical research. Software outputs should be considered in all aspects of academic performance appraisals, career applications, and promotion and tenure review cases. }
  \nrec{Workforce}{University }{Medium} {wcB}{Partnerships to support data science staff}{Universities should explore means of providing support for data-science faculty and staff, perhaps sharing capacity between academic groups or departments internally or partnerships outside the university.}
  \nrec{Workforce}{Agency }{Medium} {wcFund}{Support long-term technical capacity}{Funding agencies should explore longer-term grants aimed at building and supporting professional (non-student) data science capacity.}

\subsection{Elevating the role of software as a product of the research enterprise} \label{sec:iwft:cookies}

The ecosystem around the publishing of scholarly work has not been set up to properly account for contributions to scientific discoveries made through tools, services and other infrastructure.  Changes for the modern way in which science is done need to be made.  Publications, like ApJ and  AJ , are run by the  AAS , a professional society, and are answerable to their boards that are elected by and comprise the membership of professional researchers, who also publish in them. Therefore, it is important to educate the larger community on changes that need to be made to support modern recognition standards for  software services and then advocate for these changes with professional societies.

Social and cultural issues within the field also must be changed to normalize the appropriate acknowledgment of those who write  software and support other science infrastructure tools.
We need academics in positions of power (e.g. on promotion and tenure review committees, recruitment teams, grant review panels) to value  software as an important product of research. Although change takes time, it is important that we begin making those changes with concrete and practical suggestions that can be incrementally introduced into accepted procedures and communal norms. These suggestions include the identification of metrics that support proper assessment of the impact of  software on achieving scientific results.  In recent years, substantial improvements to enable the citation of  software and tracking of these citations has been made in astronomy and astrophysics.

%

\subsubsection{Measuring/citing the impact of software}

One key factor for improving the recognition of software within academia is to enable native software citation, that is, make it possible and required for authors to cite the software packages they have used in the process of carrying out their research, and to then count these citations in tools such as the Astrophysics Data System ( ADS ). Enabling software citation is both a technical challenge and a cultural one: recommendations for what software should be cited and when to cite it have been explored in community-wide effort at  FORCE11  \citep{Smith2016}, and follow-on efforts are exploring some of the more technical aspects of how to implement these recommendations (\cite{force11softwareimplentation}).

Within astronomy and astrophysics, the Asclepias project \footnote{\url{http://adsabs.github.io/blog/asclepias}}  --- a collaboration between  AAS publishing,  ADS, and the Zenodo data archive  \citep{henneken_edwin_2017_1011088} --- is working to enable first-class support for  software citation in  AAS journals as well as support for indexing (counting) these citations within  ADS. While this project is currently scoped to  AAS  journals only, the changes being made to support the citation and indexing of  software serve as an example for other journals to follow suit.

\subsubsection{Strategies for elevating the role of software} \label{sec:iwft:credit}

Part of the challenge of elevating the role of  software within academia is to find actionable changes that improve the career prospects of those individuals writing research  software. In this section, we outline a number of possible approaches.

\textbf{Software papers}: One approach gaining traction across a number of research disciplines is to allow papers about  software to be published in ``conventional'' journals alongside other research papers, thereby making  software more visible to the academic community, and giving  software engineers a citable ``creditable'' entity (a paper) to include on their resume. Examples of journals within astronomy that demonstrate a willingness to follow this approach include PASP \footnote{\url{https://iopscience.iop.org/journal/1538-3873}}  and  AAS  publishing, which recently changed its editorial policies to explicitly allow  software papers in their publications \citep{aassoftware}. More recently  AAS  publishing has announced a partnership with another journal specializing in  software review \citep{Vishniac_2018}.

\textbf{Enabling support for  software citation and indexing}: Another key factor in raising the visibility of research  software is to enable  software citation, count these citations, and then make these metrics visible to the world. As part of the work of the Asclepias project,  software citations are not only being counted in the astronomical literature, they are also being made visible on the  ADS  website next to the paper record on  ADS .

\textbf{Inform and educate the community about  software contributions}: Organizations play a critical role in improving the career prospects of those writing research  software as they are responsible for hiring these individuals, evaluating their performance, and making decisions about possible promotions/career advancement. One immediately actionable approach is to encourage prospective employees and current staff to list  software they have developed on their resumes and performance appraisals. This would allow review committees to include  software as part of their evaluations.

\textbf{Community prizes}:  AAS  has a collection of prizes for scientific merit, instrumentation, education, and service to the field. As it is an important part of scientific discovery,  software contributions that have had a lasting positive impact on the field, should also be recognized with a new dedicated prize and/or as a recognized example of merit within these other prize categories.

\textbf{Grants}: The amount of research funding secured is an established metric for evaluating an individual. As recommended in software, allowing existing funding streams to be utilized for  software development provides a simple mechanism for funding research  software, but also signaling community recognition for the impact and relevance of the individual writing this  software. Furthermore, widespread availability of grant funding in support of  software development would provide a strong incentive for universities to hire technical astronomers into tenure track positions.

 \nrec{Workforce}{Astronomer }{Short} {wc8}{Adopt best practices for  software citation}{Journals and reviewers should adopt best practices for assuring that  software and other science support infrastructure is properly referenced and cited in articles.  Referees and other reviewers should be trained to recognize when such acknowledgement is necessary and ask authors to provide that information.}
 \nrec{Workforce}{Manager }{Long} {wc9}{Adopt promotion metrics that acknowledge  software and other science support}{Departments and other members of the community should adopt and use suggested metrics for promotion and tenure reviews of those scientists whose work and contributions involve  software and science infrastructure.}
 \nrec{Workforce}{Agency }{Short} {wc10}{Community prizes for  software contributions}{Professional astronomy societies should create dedicated  prizes and allow for software contributions to be recognized as a criteria of merit within existing prizes. }

\section{A need for dedicated education and outreach expertise
\Contact{Amanda Bauer}}
\Contributors{Amanda E. Bauer <\mail{abauer@lsst.org}>, Britt Lundgren, Meg Schwamb, Brian Nord, Dara J Norman}
\label{sec:epo}

\bigskip
\noindent\textbf{Note:}
If you have come directly to this chapter we suggest you please read at least the Introduction in \secref{sec:intro} before delving further.
\newline


We need to capitalize on positive trends in digital literacy, the increasing use of mobile devices, and a discovery space driven by social media, through the progressive development of online resources in astronomy education and public outreach (\gls{EPO}). The goal for this chapter is to clarify and bolster the multitude of opportunities that exist to develop newly accessible online tools to engage fellow citizens in the era of petabyte-scale astronomy.

Maintaining support for astronomy research relies on our ability to effectively communicate our science and cultivate public excitement and engagement. Historically, strategic programming for astronomy \gls{EPO} in science projects has been an afterthought: the work has primarily been undertaken by astronomers who are passionate about \gls{EPO} but may lack the specific professional skills required to do it effectively at scale.  Moreover, most astronomers are not compensated for their time or rewarded by their efforts in \gls{EPO}. To maximize the public impact of large projects in the petabyte era, we must give professional credit to astronomers who do outreach work and also dedicate resources to full-time personnel to develop, execute, and evaluate modern \gls{EPO} activities.

Traditional means of public engagement (e.g. classroom visits, online videos, public lectures and panels, etc.) have demonstrated their importance and value, and have carved a niche in the landscape of public engagement. However, we have entered a new era of technology and social interaction, which necessitates new modalities for innovative pedagogical techniques, communication, and even scientific exploration. Taking advantage of opportunities of modern technology requires putting in place the appropriate professionals to create and develop the interfaces and connections to curricula that maximize adaptability and use. For example, connecting non-experts with ever larger datasets requires educators who have astronomy domain expertise (to curate and work with datasets) as well as expertise in innovative pedagogical practices.

{\bf In this new era of engagement, \gls{EPO} teams who develop ground-breaking activities and pedagogical frameworks will have started the design process as early as possible (including during construction of new facilities) and will have drawn on a number of areas of expertise: astronomical research methods, educational theory and practice, web development and design, \gls{software} engineering, and multi-modal communication.}

In this chapter, we discuss recommendations and effective practices for advancing astronomy in society through data-driven education and outreach activities for maximizing the impact large observing facilities and data centers will provide. We begin by discussing the creation of accessible online activities (\secref{epo-online}), then identify a range of skills needed to create such activities (\secref{eposkills}), and finally, we establish the benefits of resourcing dedicated \gls{EPO} groups from the earliest stages of astronomy facility planning and including \gls{EPO} as part of the mission of projects (\secref{epogroups}).

\subsection{Create accessible online activities for the public}\label{epo-online}

\nrec{EPO}{Educator, Astronomer}{Short} {epo1}{Create accessible online Activities for the Public}{To maximize the impact of astronomy in society in the rapidly approaching petabyte and exabyte eras, we recommend that projects and data centers develop accessible web interfaces and tools that enable the public to access, explore, and analyze authentic astronomical data at unprecedented scale.}

Many good arguments have been made for enabling non-professionals and students to access and engage with authentic data and professional tools. However, in practice, the increasing complexity of interfaces to large datasets can become a barrier to access and use.

User interfaces need to be attractive and intuitive for non-specialists and usable from mobile devices and platforms commonly used in schools (such as chromebooks and tablets). Interfaces created for professionals do not necessarily work for non-specialists, because they tend to have the following characteristics: 1) offer too many options; 2) do not offer a clear path toward a learning outcome; 3) too slow,  unresponsive, or burdensome for the internet connections.  Effort should be spent on user interfaces for public audiences, and ideally, on creating introductory activities as preparation for more complicated tasks.

Surveying users to assess their needs and interests helps the content design process and continues to improve the quality of an experience for users when a program in running. User testing is a regular practice for many companies that deliver a product to the public and is a process that should be adapted within astronomy \gls{EPO} programs to ensure activities remain relevant and useable.

\subsubsection{Examples of online activities}
Several examples of existing and planned infrastructures illuminate avenues for online public engagement: below, we discuss Sloan Digital Sky Survey's (\gls{SDSS}) SkyServer, Zooniverse's \gls{Citizen Science}, \gls{NASA}'s Universe of Learning, and the \gls{EPO} program of \gls{LSST}.

For over 15 years, the \gls{SDSS} has made its vast database of imaging and spectroscopic data ($\sim$200 \gls{TB}) freely available to the world. The web-based \gls{SDSS} data browser, SkyServer\footnote{\url{http://skyserver.sdss.org}}, provides a public entry point for navigating the data online. The numerous and diverse query and analysis tools available through the SkyServer are designed to meet the needs of astronomers and non-professionals alike. The benefit to this design is that any interested student or member of the public has unrestricted access to research-grade inquiries and applications of the data.  However, the large number of available features and the technical jargon that accompany them often overwhelm non-experts, as well as professional astronomers who are external to the \gls{SDSS} collaboration.

In order to better support audiences who may be put off or overwhelmed by the professional-grade access points to the \gls{SDSS} database, the \gls{SDSS} Education and Public Outreach team developed activities with simplified query tools and smaller, curated datasets to facilitate activities for pre-college educators and students (e.g. \gls{SDSS} \textit{Voyages}\footnote{\url{http://voyages.sdss.org}}). For non-specialist audiences, these activities lower the barrier to accessing the same authentic data, while providing an introduction to concepts related to both astronomy and data structures.   For students and educators who may be interested in using the data for more advanced explorations, \gls{SDSS} Voyages provides a helpful stepping stone.  The next two sections of this chapter suggest avenues to promote this transition in other ongoing and planned astronomy projects and facilities.


Citizen science represents an example of successful use of the modern age of web connectivity by directly engaging the public in scientific research. Online citizen science enables scientists to work with the general public to perform data-sorting and analysis tasks that are difficult or impossible to automate, or that would be insurmountable for a single person or for small groups of individuals to undertake \citep{2015ARA&A..53..247M}. Highly accessible citizen science activities can advance both science and learning in the era of large astronomical datasets. Moreover, most participants from the public claim that the main reason they participate is the contribution they are making to fundamental science research \citep{2017AAS...22941107C}. Through online citizen science portals such as the Zooniverse\footnote{\url{http://www.zooniverse.org}}  \citep{2011MNRAS.410..166L}, millions of volunteers have participated directly in this collaborative research experience, contributing to over 70 astronomy-based research papers. Another reason for the continued success of the Zooniverse platform in particular, is that it looks good and feels modern, even after a decade of activity. While professional astronomers are the \gls{PI}'s of citizen science projects, Zooniverse employs 13 developers, one designer, and two postdocs to lead the infrastructure development of the platform between the Adler and Oxford locations.

Members of the Zooniverse team have furthered the project's educational impact by developing a college-level data science curriculum around their crowd-sourced data. The NSF-funded Improving Undergraduate \gls{STEM} Education (\gls{IUSE}) Project: ``Engaging Introductory Astronomy Students in Authentic Research through Citizen Science" (PI: L. Trouille) is a particularly successful example of scoping big-data astronomy for a college non-major audience. This innovative curriculum equips students with the essential tools to explore the intrinsic and environmental properties of 20,000 low-redshift \gls{SDSS} galaxies that have morphological classifications from Galaxy Zoo.  This project utilizes a curated dataset in Google Sheets and a simple, plug-in tool that enables intuitive data cropping and visualization. Instead of learning about galaxies and cosmology through traditional readings and lectures, students are challenged to discover key patterns and properties of the universe themselves, through first-hand explorations of authentic astronomical data. In the process, they gain skills in quantitative analysis, improve their overall data literacy, and practice science communication. The curriculum specifically provides an opportunity to discuss the complications and limitations of authentic data, and the challenges of framing a question that can be effectively tested with the data one has in hand. The project is a great case study of delivering specific, high-impact learning outcomes through an analysis of authentic data, without requiring students to navigate full-scale datasets or jargon-rich professional tools for visualization and analysis.

NASA's Universe of Learning\footnote{\url{https://www.universe-of-learning.org/}} offers a variety of individual online web pages that are well presented. A potential challenge for a typical user who finds one of these pages is knowing what to do next. Beyond exploring the beautiful multi-wavelength images space telescopes provide, there is not a clear path for a user to navigate toward specific learning outcomes or experiences.



LSST’s \gls{EPO} program\footnote{https://www.lsst.org/about/epo} is unique among ground-based telescope projects: not only is it being constructed in tandem with the physical observatory itself, but the outreach program is funded at 2\% of the project cost. \gls{EPO} products will go live when the \gls{LSST} Survey begins in 2022. \gls{EPO} products were included from the beginning as part of the construction Project deliverables, because they faced similarly unique challenges as the data resulting from the survey itself. During its design phase, the \gls{EPO} team selected specific audiences and invested in user needs assessments to examine what these audiences want, and cannot find elsewhere. Some major findings include the necessity for mobile-friendly interfaces, a clear path toward learning goals, and educators needing no new \gls{software} to download in order to introduce classroom activities.  This has shaped the overall strategy for \gls{LSST} \gls{EPO} development and the skill sets needed on the \gls{EPO} Team, which is a small, interdisciplinary team of astronomers, writers, designers, educators, and developers. The mission of \gls{LSST} \gls{EPO} is ``to offer accessible and engaging online experiences that provide non-specialists access to, and context for, \gls{LSST} data so anyone can explore the Universe and be part of the discovery process.''

The operations website will feature news about \gls{LSST} discoveries, profiles of \gls{LSST} scientists and engineers and their work, and will be optimized for use on mobile devices. The \gls{EPO} team is also developing online, data-driven classroom investigation activities for students in advanced middle school through college. The topics cover commonly-taught principles in astronomy and physics, and each investigation is designed for use with Next Generation Science Standards (\gls{NGSS}) in the United States and the Curriculum Nacional in Chile. All investigations come with support and assessment materials for instructors and no special \gls{software} is needed to access the investigations, which will be available in English and Spanish. \gls{LSST} \gls{EPO} will maintain an easy-to-use gallery of high-quality multimedia visualizations that can be downloaded and integrated into exhibits and presentations. Finally, \gls{LSST} \gls{EPO} will provide support to researchers who create \gls{Citizen Science} projects using \gls{LSST} data, including a dedicated project-building tool on the Zooniverse platform. The infrastructure to host these activities is being built during construction and will take several years. Another critical task during construction is building prototypes and performing user testing, which has continually proven to improve the user experience and usability of interfaces.

A consistent theme that emerges when examining these examples is that well-defined learning outcomes for activities, curated access to authentic data, and simple, intuitive design are important to prepare for the \gls{EPO} response to the large data we will collect in the 2020s. The remaining sections identify areas of expertise \gls{EPO} teams can employ to achieve these outcomes.

\subsection{Bring expertise to astronomy education and outreach teams}\label{eposkills}

\nrec{EPO}{Manager, Educator}{Medium}{epo3}{Bring dedicated experts onto astronomy education and outreach teams}{To create the accessible online interfaces that maximize public impact in the next decade, we recommend supporting dedicated education and outreach teams that pair astronomers with technical and education specialists to increase relevance, adoptability, and accessibility of activities.}

The large-scale data challenges that face astronomy described in this paper also represent challenges and opportunities for formal education, public outreach, and science communication.  A natural instinct for astronomers may be to adapt their new computational experience to outreach efforts. This is a noble goal, but astronomers are not be expected to know effective practices around mobile-friendly development, intuitive user interfaces for the public, marketing through social media, or how to connect astronomy activities to formal education curriculum standards.  A team of \gls{EPO} experts can advise and assist with these areas, which are essential to build successful activities that are discoverable and adoptable.

We recommend astronomy organizations support creating \gls{EPO} teams with expertise in relevant areas. It is understood that to reach maximal impact of outreach activities, these individuals work with astronomers to combine astronomy and data science expertise with specific \gls{EPO} expertise and experience. This section describes options for areas of expertise and roles that can be brought on to achieve specific goals.

Educators in the \gls{US} are currently required to submit paperwork to demonstrate that they are teaching specific topics related to curriculum standards.  An \gls{EPO} {\bf education specialist or instructional designer} brings knowledge of relevant curriculum standards and rubrics (for example, the Next Generation Science Standards\footnote{NGSS: \url{https://www.nextgenscience.org/}}) and is able to connect astronomy activities to topics educators must cover. This is the most relevant for K-12 formal education in a traditional setting or homeschooling. An educational specialist can build professional development programs to increase confidence for bringing such activities into their classrooms if there is not an expert available to join in person.

An education specialist can also tap into educator networks to advertise existing programs and perform professional development. An example is the National Science Teachers Association (\gls{NSTA}) annual meeting and \gls{AAS}. An education specialist working with an astronomer can create curate datasets to achieve specific learning outcomes without overwhelming non-specialists \citep{Rebull18}.

An {\bf Information Officer or Communications Manager} can act as a primary contact for a facility and also set overall communication strategies and implementation plans. Topics covered in such strategies could include audiences, content priorities, communication channels, messaging, procedures, and more.

An {\bf Outreach Specialist} could serve a range of purposes depending on the needs of the group.  This could be a science writer, someone who responds to and directs questions received from audience groups, or contributes to social media presence.  or an astronomer trained in science communication. If this person has astronomy training, he/she could work with astronomical datasets to curate options for the public.

Social Media is becoming increasingly important as a source of news and information in society. Dedicating a full-time equivalent (or more) to the role of {\bf social media engagement specialist} increases awareness of \gls{EPO} activities and engages various audiences to participate with activities that exist.

Overall branding, the look and feel of online activities, and developing interesting graphics and images to support press releases or other activities are the role of a {\bf Graphic Designer}.

An {\bf evaluation specialist} informs methods for understanding the impact of programs on specific audience groups. The most benefit occurs when the method for evaluating the success of a program is built into the development of the program itself. Metrics could include and are not limited to web analytics, short or long surveys, interviews, login requests, focus groups, web submission forms, and social media interactions.

A {\bf web developer} considers the user interface and experience when visiting a site. Mobile-friendly accessibility is a requirement for non-specialists since most users of an online interface will discover the materials via social media and will access them from a mobile device, not a desktop  platform. In addition, the most common machines used by schools are chromebooks and potentially weak internet connections, which require lightweight design. Development needs to satisfy these requirements are best implemented by experts in the field.

A {\bf Software Architect} designs, deploys, and maintains production services for an online program. It is important to not overburden internet systems that can be common in classroom settings or non-urban areas.

A {\bf Project Manager} oversees the detailed budget, schedule, contracts, documentation, and reporting.  This role is important for programs being built during the construction of an astronomical facility.

\subsection{Fund dedicated astronomy education and outreach groups}\label{epogroups}

\nrec{EPO}{Agency}{Long}{epo2}{Fund dedicated or centralized astronomy education and outreach groups}{We recommend that funding agencies supporting the development and operation of large astronomical observing and data facilities fund professional education and outreach groups who can provide strategy, oversight, and best practices to maximize the impact of outreach efforts, and encourage \gls{EPO} efforts to be part of a project's mission.}

Having a dedicated individual or team to develop the \gls{EPO} program for a specific facility can improve efficiency, impact, and cost effectiveness. Strategic planning provides an opportunity to emphasize the identity of a particular large facility; to identify non-specialist audiences who could benefit the most from dedicated engagement; put into place best practices in outreach and communication programs; and complement the overall landscape of astronomy \gls{EPO} efforts. It is important that the \gls{EPO} professionals are employed directly at professional telescope facilities in order to emphasize the uniqueness of the program, build and maintain relationships with those doing the technical and scientific work, and help handle the astronomy-specific data products that currently require a reasonable level of understanding to interpret and use (see \secref{sec:data}).

A dedicated \gls{EPO} team also serves as a resource for enabling astronomers working with large datasets and data facilities to do more impactful and wide-reaching outreach. Groups that are specifically charged to do \gls{EPO} can improve the impact of the existing \gls{NSF} Broader Impacts investment by supporting astronomers to tap into existing programs. This improves discoverability of the \gls{EPO} work astronomers are doing increases the likelihood of achieving Broader Impact goals at both the individual and \gls{NSF} levels.

An \gls{EPO} team could provide any of the following benefits:
\begin{itemize}
\item Conducting science communication and media training sessions for astronomers doing these activities.
\item Providing introductions to various social media platforms that can be used for unique outreach experiences.
\item Marketing and promoting activities through established social media and common online training resources (e.g. Code Academy).
\item Creating or tapping into a centralized repository for people looking for resources
\item Creating opportunities for collaboration between astronomers and existing outreach infrastructure that will promote success and provide wide-reaching impact. Examples include {\em Journey Through the Universe} in Hawai’i or {\em AstroDay} in Chile, both led by Gemini \gls{EPO}.
\item Performing user needs assessments and user testing to improve the quality of existing activities and to develop new programs that meet the needs of specific audiences.
\item Evaluating and reporting on the impact of \gls{EPO} activities. Evaluation methods can and should be built into program design.
\item Providing guidance for astronomers when developing science drivers and use cases for educational materials and public interfaces related to their research expertise.
\item Broadening participation to non-traditional audience groups.
\end{itemize}




The timing for building expert \gls{EPO} teams should occur during the construction of a new facility and be included as part of the project's mission.  Starting early affords time to implement appropriate strategy and infrastructure.  Educational materials, supplemental professional development materials, striking visualizations and images and communications strategies should be ready at the start of a project to maximize the public impact of the facility.

In this chapter, we discussed recommendations and effective practices that can be employed to maximize the impact of large astronomy facilities and data centers in the next decade. We prefaced the need for creating accessible online astronomy activities for the public and identified a range of skills needed to create such activities. Finally, we established the benefits of resourcing dedicated \gls{EPO} groups from the earliest stages of astronomy facility planning and even including \gls{EPO} as part of the mission of projects.

These recommendations are based on two main things: trends seen elsewhere on the web that successfully respond to this new era of technology and social interactions on the web, and case studies within astronomy that demonstrate appropriate avenues for increasing engagement and accessibility through online activities.

Note: This chapter is the basis for an Astro2020 Decadal Survey \gls{APC} white paper which will go into detail on resourcing and prioritizing recommendations. If you are interested in commenting or constributing, please contact Amanda E. Bauer <abauer@lsst.org>.

\section{Conclusion }
\label{sec:conclusion}

We had a good opportunity to think about a range of topics which have been detailed in this document. Several \gls{APC} white papers will be published using this  as a basis.
If you are interested in contributing to or endorsing \href{white papers on these topics sign up here}{https://tinyurl.com/y2ksemp2}\footnote{ \url{https://tinyurl.com/y2ksemp2}}  or contact the authors listed in this document.
We do not intend this to be the solution to all issues rather a discussion of potential ways forward for the next decade.
We shall host a \href{https://petabytestoscience.github.io/}{third and final workshop}\footnote{\url{https://petabytestoscience.github.io/}} in October 2019 which will explore practical approaches to dealing with some of the recommendations raised here.

\subsection*{Acknowledgments}

The workshops in this series were supported by The Kavli Foundation\footnote{\url{https://www.kavlifoundation.org/}}. The Kavli Foundation is dedicated to advancing science for the benefit of humanity, promoting public understanding of scientific research, and supporting scientists and their work.

\printnoidxglossaries


\bibliographystyle{yahapj}
\bibliography{main}

\end{document}